\documentclass[%
 article,two column,
superscriptaddress,
 amsmath,amssymb,
 aps,
]{revtex4-1}
\usepackage{graphicx}
\usepackage{dcolumn}
\usepackage{bm}
\usepackage{bbold}
\usepackage{calc}
\usepackage{changes}
\usepackage{xcolor}
\usepackage{hyperref}
\usepackage{url}

\usepackage{color}
\usepackage{xcolor,colortbl}
\usepackage{comment}
\newcommand {\nc} {\newcommand}
\nc {\IR} [1]{\textcolor{red}{#1}}
\nc {\IB} [1]{\textcolor{blue}{#1}}
\nc {\IM} [1]{\textcolor{magenta}{#1}}

\hypersetup{
    colorlinks=true,
    linkcolor=blue,
    citecolor=blue,
    filecolor=magenta,      
    urlcolor=blue,
    pdftitle={Overleaf Example},
    pdfpagemode=FullScreen,
    }

\begin{document}


\title{Chiral EFT calculation of neutrino reactions in warm neutron-rich matter}
\author{Eunkyoung Shin}%
\email{shinek@tamu.edu}
\affiliation{%
Cyclotron Institute, Texas A\&M University, College Station, TX 77843, USA\\
Department of Physics and Astronomy, Texas A\&M University, College Station, TX 77843, USA}%

\author{Ermal Rrapaj}
\email{ermalrrapaj@gmail.com}
\affiliation{NERSC, Lawrence Berkeley National Laboratory, Berkeley, California 94720, USA\\
University of California, Berkeley, CA 94720, USA\\
RIKEN iTHEMS, Wako, Saitama 351-0198, Japan}%

\author{Jeremy W. Holt}
\email{holt@comp.tamu.edu}
\affiliation{%
Cyclotron Institute, Texas A\&M University, College Station, TX 77843, USA\\
Department of Physics and Astronomy, Texas A\&M University, College Station, TX 77843, USA}

\author{Sanjay K. Reddy}
\email{sareddy@uw.edu}
\affiliation{
Institute for Nuclear Theory, University of Washington, Seattle, WA 98195, USA
}%

\date{\today}

\begin{abstract}
Neutrino scattering and absorption rates of relevance to supernovae and neutron star mergers are obtained from nuclear matter dynamical structure functions that encode many-body effects from nuclear mean fields and correlations. We employ nuclear interactions from chiral effective field theory to calculate the density, spin, isospin, and spin-isospin response functions of warm beta-equilibrium nuclear matter. We include corrections to the single-particle energies in the mean field approximation as well as vertex corrections resummed in the random phase approximation (RPA), including, for the first time,  both direct and exchange diagrams. 
We find that correlations included through the RPA redistribute the strength of the response to higher energy for neutrino absorption and lower energy for antineutrino absorption. This tends to suppress the absorption rate of electron neutrinos across all relevant energy scales. In contrast, the inclusion of RPA correlations enhances the electron antineutrino absorption rate at low energy and supresses the rate at high energy. These effects are especially important at high-density and in the vicinity of the neutrino decoupling region. Implications for heavy element nucleosynthesis, electromagnetic signatures of compact object mergers, supernova dynamics, and neutrino detection from galactic supernovae are discussed briefly.   

\end{abstract}

\maketitle

\section{\label{sec:level1}Introduction}
 
Neutrinos dominate energy, momentum, and lepton number transport in extreme astrophysical phenomena. Their scattering and absorption rates in hot and dense nuclear matter play a critical role in core-collapse supernovae \cite{burrows99,oconnor15,melson15,roberts16}, proto-neutron star cooling \cite{pons98,roberts17}, and neutron star mergers \cite{sekiguchi10,wanajo14,endrizzi20,sumiyoshi21,cusinato21}. The effects of nuclear mean fields and correlations on neutrino reaction rates are encoded in dynamical structure functions related to the imaginary part of nuclear response functions. In the past, nuclear matter response functions have been studied using a variety of nuclear interactions and many-body approximations, including nonrelativistic and relativistic mean field models \cite{sawyer89,reddy98,reddy99,martinez-pinedo12,roberts12,pastore12}, Fermi liquid theory \cite{iwamoto82,burrows98,burrows99}, the virial expansion \cite{horowitz06,horowitz17,bedaque18}, and pseudopotentials \cite{rrapaj15,bedaque18}. Recent studies \cite{martinez-pinedo12,roberts12,rrapaj15} have highlighted the important role of nuclear mean fields for calculating charged-current reactions, such as neutrino and anti-neutrino absorption, in the supernova neutrinosphere. Here the large asymmetry between proton and neutron densities leads to a strong splitting of the proton and neutron mean fields that enhance neutrino absorption and suppresses anti-neutrino absorption. This, in turn, affects the composition of matter ejected from supernovae and neutron star mergers as well as neutrino flavor and energy distributions that terrestrial neutrino detectors may observe.

In addition, the use of high-precision nucleon-nucleon interactions \cite{bacca12,rrapaj15,bartl16} to calculate neutrino scattering and reaction cross sections in hot and dense matter has illustrated the important role of contributions beyond one-pion-exchange together with nonperturbative resummations of the nucleon-nucleon interaction in the particle-particle channel (see also Ref.~\cite{Hanhart:2000ae} for the role of these effects on neutrino production).  At zero energy transfer, the momentum dependence of the static density response function of neutron-rich matter is related to the poorly known isovector gradient contribution in nuclear energy density functionals used to model neutron star crusts \cite{lim17,carreau19}. Recent quantum Monte Carlo simulations of neutron matter using high-precision nuclear forces have provided constraints on the isovector gradient contribution through studies of neutron drops \cite{gandolfi11} and the static density response function \cite{buraczynski16}.

\begin{figure*}[t]
\centering
\includegraphics[width=\textwidth]{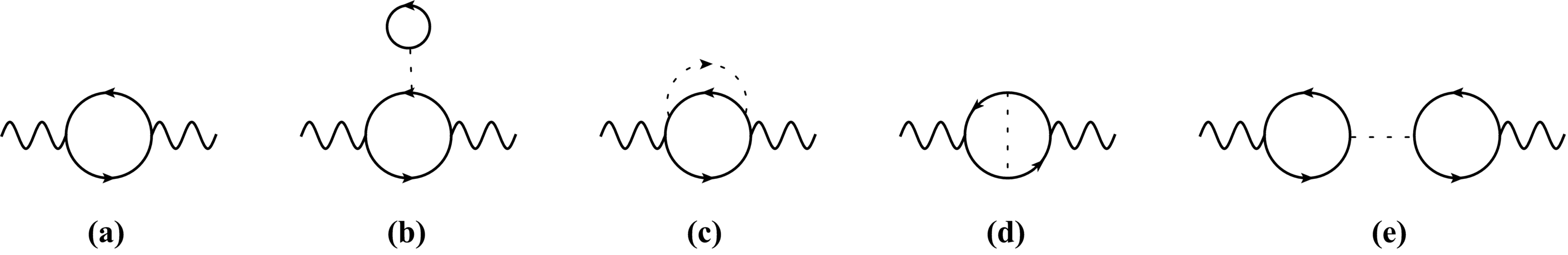}
\caption{Zeroth-order and first-order diagrammatic contributions to the nuclear response function. The zeroth-order diagram is labeled (a), the first-order direct and exchange mean field contributions are labeled (b) and (c), and the first-order direct and exchange vertex corrections are labeled (d) and (e). Solid lines represent nucleons, wavy lines represent the coupling to an external probe, and dashed lines represent the nucleon-nucleon interaction.}
\label{diagrams}
\end{figure*}

In this study, we employ nuclear forces based on chiral effective field theory to investigate beyond-mean-field corrections to spin and density response functions of nuclear matter under ambient conditions typical of supernova and neutron star merger neutrinospheres, where the nucleon number density varies in the range $n=10^{11}-10^{13}$\,g/cm$^3$ and the temperature varies in the range $T=5-10$\,MeV. Specifically, we calculate Hartree-Fock mean field corrections as well as resummed particle-hole vertex corrections in the random phase approximation (RPA). In contrast to naive order-by-order perturbation theory, the RPA with self-consistent Hartree-Fock mean fields provides a thermodynamically consistent ``conserving approximation'' \cite{baym61,harris78} for which the computed dynamic structure function is guaranteed to respect sum rules, detailed balance, and positivity for all values of the energy and momentum transfer.
We find that at the densities of relevance, RPA vertex corrections can be as important as the mean field effects studied in earlier work and should be included in a consistent description of nuclear matter response functions for astrophysical applications.
The paper is organized as follows. In Section \ref{meth}, we present expressions for density and spin response function in isospin-asymmetric nuclear matter at finite temperature with mean field and RPA vertex corrections. In particular, we outline an exact matrix inversion method \cite{harris78} for calculating RPA response functions, including both direct and exchange terms for an arbitrary nucleon-nucleon potential. In Section \ref{res} we calculate neutral-current and charged-current density and spin response functions for a range of thermodynamic conditions in beta-equilibrium matter. Finally, we conclude with a summary and outlook in Section \ref{con}.

\section{Response functions in isospin-asymmetric nuclear matter}
\label{meth}

In the present section we derive expressions for the first-order mean field and vertex corrections to the response functions of homogeneous nuclear matter at nonzero temperature. We also outline the calculation of the RPA response function employing a matrix eigenvalue method. In the region of the supernova or neutron star merger neutrinospheres, where neutrinos decouple from nuclear matter and their free-streaming energy spectrum is set, the proton fraction is small $Y_p \approx 0.05-0.10$, the temperature is warm $T = 5 - 10$\, MeV, and the matter is dilute $\rho \approx 10^{11}-10^{13}$\,g/cm$^3$. We assume beta equilibrium, which provides the restriction
\allowdisplaybreaks
\begin{equation}
\label{beta}
\mu_n - \mu_p - \mu_e = 0
\end{equation}
on the proton and neutron chemical potentials and hence their number densities
\begin{equation}
n_i = \frac{2}{(2\pi)^3} \int d^3 k \frac{1}{1+e^{(e_i(k)-\mu_i)/T} },
\label{dens}
\end{equation}
for $i=\{n,p,e\}$. Together with charge neutrality
$\rho_e = \rho_p$, the above equations must be solved self consistently to determine the proton, neutron, and electron chemical potentials for a fixed baryon number density and temperature.

In the mean field approximation, one also has to calculate the nucleon energy-momentum dispersion relation for protons $e_p(k) = k^2 / (2M_p) + \Sigma_p(k)$, and likewise for neutrons, where $\Sigma_p(k)$ is the self-energy. The Hartree-Fock conserving approximation \cite{harris78} consists of computing the irreducible self-energy to first order in perturbation theory and the response function to all orders in the random phase approximation. The resulting perturbative approximation to the response function is then guaranteed to satisfy properties of the exact response function, such as sum rules and strictly positive dynamical structure functions.

In the present work we employ nucleon-nucleon (NN) potentials derived from chiral effective field theory (ChEFT) \cite{entem03,coraggio07,marji13}, which provides a systematic expansion of nuclear two and many-body forces. In ChEFT the long-range part of the nuclear force comes from pion-exchange processes constrained by chiral symmetry, while the short-range part is encoded in a set of contact terms fitted to nucleon-nucleon scattering and deuteron properties. The high-momentum components of the ChEFT nuclear potentials employed in this study are regulated by exponential functions with a characteristic momentum scale $\Lambda$ and smoothness parameter $n$:
\begin{equation}
    f(p,p') = \exp[-(p'/\Lambda)^{2n}-(p/\Lambda)^{2n}],
\end{equation}
where $\vec p$ and $\vec p^{\,\prime}$ are the incoming and outgoing relative momenta of the two nucleons. We employ ChEFT potentials with cutoff scales $\Lambda = 414, 450, 500$\,MeV and associated smoothness parameters $n=10, 3, 2$ respectively. When supplemented by the ChEFT three-body force that appears at next-to-next-to-leading order (N2LO) in the chiral expansion, this set of nuclear potentials has been shown to predict well the properties of nuclear matter, such as the equation of state \cite{Holt,Wellenhofer:2014hya}, optical potential \cite{Whitehead21}, and quasiparticle interaction in Fermi liquid theory \cite{Holt:2011yj}. In dilute neutron-rich matter, three-body forces give negligible contribution to single-particle and response properties of the medium. We therefore neglect three-body forces in the present work.

\subsection{Nuclear matter response functions at 0th order in perturbation theory}

In Figure \ref{diagrams} we show the zeroth-order and first-order perturbation theory contributions to nuclear response functions. The wavy lines denote the coupling to a $W$ or $Z$ boson defined in the non-relativistic limit by its vector/axial vector and isoscalar/isovector nature. The dashed lines in Figure \ref{diagrams} represent the nucleon-nucleon interaction.

The zeroth-order contribution to the neutral-current density response function $\chi_{\rho}$, the neutral-current spin response function $\chi_{\sigma}$, the charged-current density response function $\chi_{\tau \rho}$ for electron-neutrino absorption, and the charged-current spin response function for electron-neutrino absorption $\chi_{\tau \sigma}$ in isospin-asymmetric nuclear matter at nonzero temperature are given by
\begin{widetext}
\begin{align}
\label{res0}
\chi^{(0)}_{\rho}(\vec{q},\omega) &= \sum_{s_1s_2t_1t_2} \! \int \! \frac{d\vec{k}}{(2\pi)^3} \frac{f_{\vec k, t_1} - f_{\vec k + \vec q,t_2}}{ \omega+e_{\vec k, t_1}-e_{\vec{k}+\vec{q}, t_2} + i \eta}\delta_{s_1,s_2}\delta_{t_1,t_2}, \\ \nonumber
\chi^{(0)}_{\sigma}(\vec{q},\omega) &= \sum_{s_1s_2t_1t_2} \! \int \! \frac{d\vec{k}}{(2\pi)^3} \frac{f_{\vec k, t_1} - f_{\vec k + \vec q,t_2}}{ \omega+e_{\vec k, t_1}-e_{\vec{k}+\vec{q}, t_2} + i \eta} |\langle s_1 | \sigma_z | s_2 \rangle|^2 \delta_{t_1,t_2}, \\ \nonumber\chi^{(0)}_{\tau \rho}(\vec{q},\omega) &= \sum_{s_1s_2} \! \int \! \frac{d\vec{k}}{(2\pi)^3} \frac{f_{\vec k, n} - f_{\vec k + \vec q,p}}{ \omega+e_{\vec k, n}-e_{\vec{k}+\vec{q}, p} + i \eta}\delta_{s_1,s_2}, \\
\nonumber\chi^{(0)}_{\tau \sigma}(\vec{q},\omega) &= \sum_{s_1s_2} \! \int \! \frac{d\vec{k}}{(2\pi)^3} \frac{f_{\vec k, n} - f_{\vec k + \vec q,p}}{ \omega+e_{\vec k, n}-e_{\vec{k}+\vec{q}, p} + i \eta} |\langle s_1 | \sigma_z | s_2 \rangle|^2,
\end{align}
\end{widetext}
where the sums are over the single-particle spin projections $s_1, s_2$ and isospin projections $t_1,t_2$ of the particle-hole pair, $f_{\vec k, t}$ is the Fermi-Dirac distribution function for a nucleon with momentum $\vec k$ and isospin projection $t$, and $e_{\vec k, t}$ is the single-particle energy for a nucleon with momentum $\vec k$ and isospin projection $t$.

In the case of noninteracting protons and neutrons, the density and spin response functions in the neutral-current or charged-current channels are identical since $\sum_{s_1 s_2} \delta_{s_1,s_2} = \sum_{s_1 s_2} |\langle s_1 | \sigma_z | s_2 \rangle|^2 = 2$. In addition, the single-particle energies in Eq.\ \eqref{res0} are simply the free-space kinetic energies $e_{\vec k}=k^2/(2M)$. Including effects from momentum-dependent mean fields for protons $\Sigma_p(k)$ and neutrons $\Sigma_n(k)$, the single-particle energies that enter in the Fermi-Dirac distribution functions and the energy denominators of Eq.\ \eqref{res0} become
\begin{equation}
e_{\vec k,n} = \frac{k^2}{2M}+\Sigma_n(k), \hspace{.2in} e_{\vec k,p} = \frac{k^2}{2M}+\Sigma_p(k).
\end{equation}
At the Hartree-Fock level (first-order perturbation theory), the proton and neutron single-particle potentials can be well described by the effective mass approximation
\begin{equation}
e_{\vec k,n} = \frac{k^2}{2M_n^*}+U_n, \hspace{.2in} e_{\vec k,p} = \frac{k^2}{2M^*_p}+U_p.
\end{equation}
In the case that $M^*_{n,p}\simeq M$, the mean field strengths $U_n$ and $U_p$ modify only the energy denominators in Eq.\ \eqref{res0}, which results in a shift of the imaginary part of the response functions. Although the mean field energies $e_{\vec k,n}$ and $e_{\vec k,p}$ also enter in the definition of the Fermi-Dirac distribution functions, a simple mean field shift will be absorbed into a redefinition of the chemical potential. The inclusion of nuclear mean fields at the Hartree-Fock level in the calculation of the $0^{\rm th}$-order response functions $\chi^{(0,{\rm MF})}(q,\omega)$ corresponds to iterating diagrams of type (b) and (c) in Figure \ref{diagrams} to all orders in perturbation theory. For neutrino scattering and absorption, the main effect will be a shift of the $\omega$-dependent imaginary response for fixed momentum transfer $\vec q$. For additional technical details regarding the calculation of nucleon single-particle potentials starting from realistic nucleon-nucleon interactions, the reader is referred to Refs.\ \cite{holt13,rrapaj15}.

\subsection{Random Phase Approximation (RPA)}
\label{sec:rpa}

\begin{figure}[t]
\centering
\includegraphics[width=\linewidth]{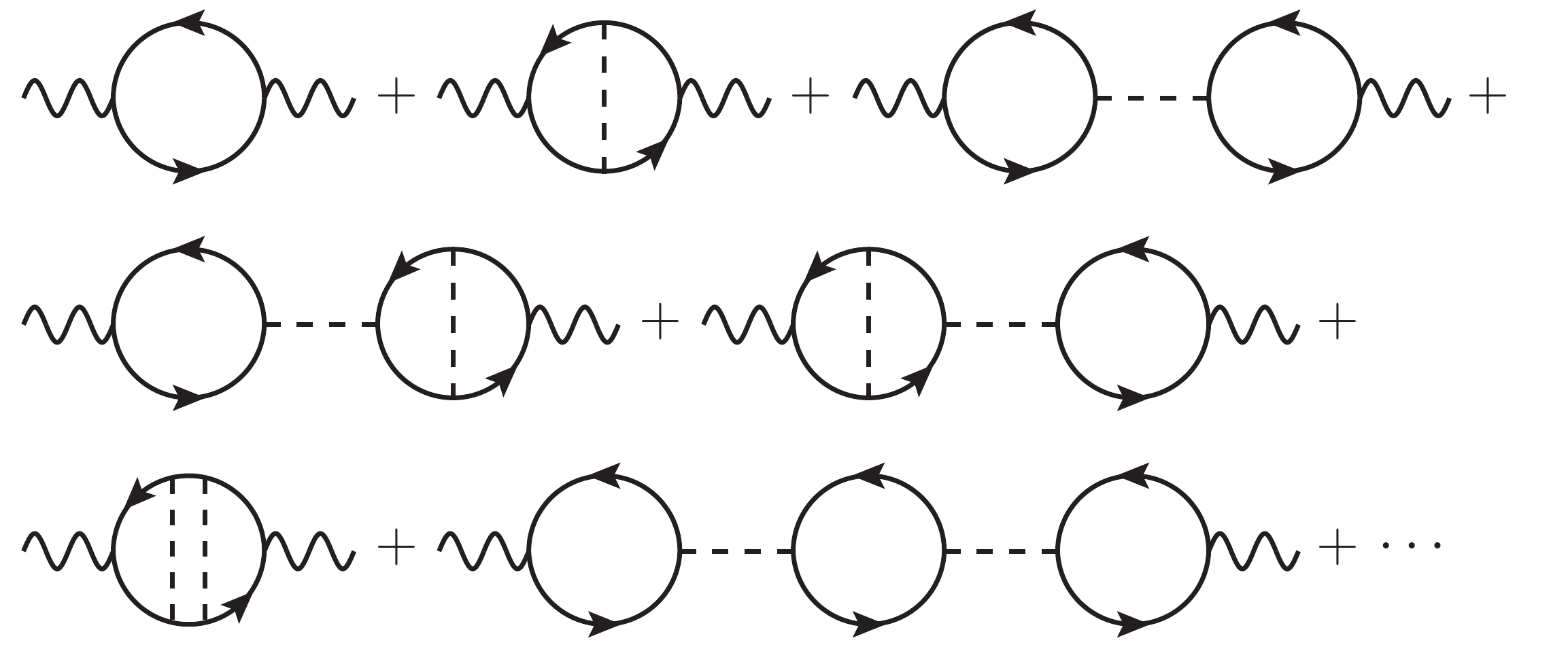}
\caption{Diagrammatic contributions to the RPA response function. Solid lines represent nucleons, wavy lines represent the coupling to an external probe, and the dashed lines represent the nucleon-nucleon interaction.}
\label{rpadiagrams}
\end{figure}

In many-body perturbation theory, an order-by-order calculation of response functions can lead to unphysical dynamic structure functions that do not satisfy constraints, such as positivity and sum rules that relate the long-wavelength response to thermodynamics. The formal solution to this problem obtained by Baym and Kadanoff \cite{baym61} is to construct so-called ``conserving approximations''  that relate the one-body and two-body propagators in such a way as to maintain conservation laws of energy, momentum, angular momentum, and particle number. In the Hartree-Fock conserving approximation, the one-body propagator is constructed at the self-consistent mean field level, while the two-body propagator resums to all orders the direct and exchange particle-hole diagrams in the RPA. The RPA leads to a linear inhomogeneous integral equation for the particle-hole vertex function. The discretized version of this equation can be re-expressed \cite{harris78} as a linear algebraic equation that can be solved through matrix inversion. In the rest of this section, we will outline the solution of the RPA response function and present several benchmarks to test the method.

\begin{figure}[t]
\includegraphics[width=\linewidth]{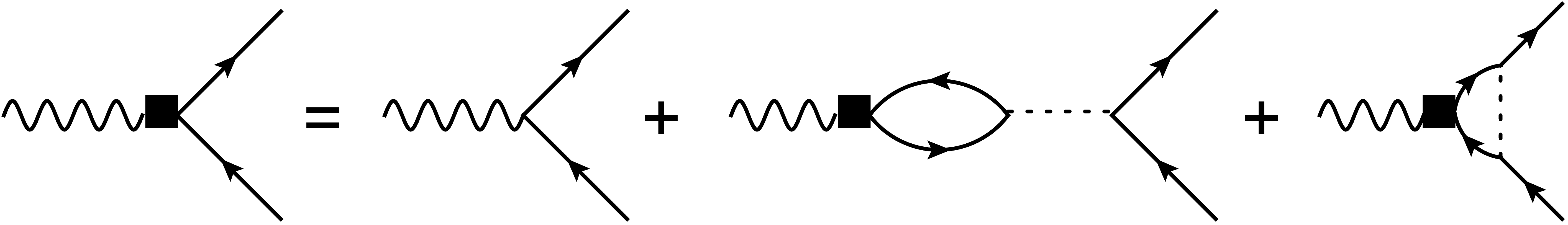}
\caption{Vertex function in the random phase approximation. The wavy line represents the weak current, the dashed line is the nucleon-nucleon interaction carrying energy and momentum, and the dark square is the dressed vertex. The solid lines represent propagating nucleons.}
\label{RPA_dia}
\end{figure}

In Figure \ref{rpadiagrams}, we show the class of response function diagrams to be resummed in the RPA, which includes all particle-hole bubble diagrams, ladder diagrams, and combinations thereof with dressed intermediate-state propagators in the mean field approximation. This infinite set of diagrammatic contributions to nuclear matter response functions cannot be resummed directly but instead must proceed through the resummed particle-hole density vertex function $L(\vec k_1 s_1;\vec q\, \omega)$. In the case of the charged-current density response function, the density vertex function satisfies
\begin{equation}
\chi_{\tau \rho}(\vec q,\omega) = \sum_{s} \int \! \frac{d\vec k}{(2\pi)^3} L(\vec k s;\vec q\, \omega).
\end{equation}
The 0th-order and 1st-order vertex functions are therefore given by
\begin{widetext}
\begin{align}
L^{(0)}(\vec k s; \vec q\, \omega) &= \frac{f_{\vec k, n} - f_{\vec k + \vec q, p}}{ \omega + e_{\vec k, n} - e_{\vec k +\vec q, p } + i \eta},
\\ \nonumber
L^{(1)}(\vec k s; \vec q\, \omega) &= \frac{f_{\vec k, n} - f_{\vec k + \vec q, p}}{ \omega+e_{\vec k, n}-e_{\vec k +\vec q, p } + i \eta}
\sum_{s^\prime} \int \! \frac{d\vec k^\prime}{(2\pi)^3} \frac{ ( f_{\vec k^\prime, n} \! - f_{\vec k^\prime + \vec q,p} ) \langle \vec k \vec k^\prime\! +\! \vec q, s s^\prime, n p | \bar V | \vec k \! +\! \vec q \, \vec k^\prime, s s^\prime, p n \rangle} { \omega + e_{\vec k^\prime, n}-e_{\vec k^\prime + \vec q, p } + i \eta }.
\end{align}
\end{widetext}
Summing the higher-order particle-hole bubble and ladder diagrams, one obtains the integral equation
\begin{eqnarray}
\label{inteq1}
&&\hspace{-.12in} L(\vec k s;\vec q\, \omega) = L_0(\vec k s;\vec q\, \omega) + L_0(\vec k s;\vec q\, \omega)
\nonumber \\ 
&&\times \sum_{s^\prime} \int \! \frac{d\vec k^\prime}{(2\pi)^3} \langle \vec k \vec k^\prime\! +\! \vec q, s s^\prime, n p | \bar V | \vec k \! +\! \vec q \, \vec k^\prime, s s^\prime, p n \rangle
\nonumber \\  
&&\times L(\vec k^\prime s^\prime;\vec q\, \omega),
\end{eqnarray}
shown diagrammatically in Figure \ref{RPA_dia}.
For a spin-saturated system, the vertex function for the density response function is independent of spin. One can then average Eq.\ \eqref{inteq1} over the spin to obtain
\begin{eqnarray}
\label{inteq2}
&&\hspace{-.12in} L(\vec k;\vec q\, \omega) = L_0(\vec k;\vec q\, \omega) + L_0(\vec k;\vec q\, \omega)
\int \! \frac{d\vec k^\prime}{(2\pi)^3} L(\vec k^\prime;\vec q\, \omega)
\nonumber \\ 
&&\times \left [ \frac{1}{2} \sum_{s s^\prime} \langle \vec k \vec k^\prime\! +\! \vec q, s s^\prime, n p | \bar V | \vec k \! +\! \vec q \, \vec k^\prime, s s^\prime, p n \rangle \right ],
\end{eqnarray}
where for convenience we will denote the quantity in squared brackets as ${\cal V}(\vec k,\vec k^\prime)$, suppressing the explicit dependence on $\vec q$. Writing the integral in Eq.\ \eqref{inteq2} as a summation over a discrete set $\{\vec k_1, \vec k_2, \dots \}$ of momentum-space mesh points with associated mesh weights $\{w_1, w_2, \dots \}$, one can rewrite Eq.\ \eqref{inteq2} as a matrix equation whose formal solution is
\begin{equation}
L = \left [ N^{-1} \left ( E + ( \omega + i \eta) \mathbb{1} \right ) - {\cal V} \right ]^{-1} B,
\label{lmat}
\end{equation}
where $L$ is a vector with elements
\begin{equation}
  L =
  \begin{bmatrix}
    L(\vec k_1;\vec q\, \omega) \\
    L(\vec k_2;\vec q\, \omega) \\
    \vdots
  \end{bmatrix},
  \label{lmatrix}
\end{equation}
$N$ is a diagonal matrix with elements
\begin{equation}
N = 
\begin{pmatrix}
f_{\vec k_1, n} - f_{\vec k_1 + \vec q, p} & 0 & \cdots \\
0 & f_{\vec k_2, n} - f_{\vec k_2 + \vec q, p} & \cdots \\
\vdots  & \vdots  & \ddots
\end{pmatrix},
\end{equation}
$E$ is a diagonal matrix with elements
\begin{equation}
E = 
\begin{pmatrix}
e_{\vec k_1, n} - e_{\vec k_1 + \vec q, p } & 0 & \cdots \\
0 & e_{\vec k_2, n} - e_{\vec k_2 + \vec q, p } & \cdots \\
\vdots  & \vdots  & \ddots
\end{pmatrix},
\end{equation}
${\cal V}$ is the matrix
\begin{equation}
{\cal V} = 
\begin{pmatrix}
w_1 {\cal V}(\vec k_1, \vec k_1) & w_2 {\cal V}(\vec k_1, \vec k_2) & \cdots \\
w_1 {\cal V}(\vec k_2, \vec k_1) & w_2 {\cal V}(\vec k_2, \vec k_2) & \cdots \\
\vdots  & \vdots  & \ddots
\end{pmatrix},
\label{vmatrix}
\end{equation}
and $B$ is a vector whose elements are all 1.

\begin{figure}[t]
\centering
\includegraphics[width=0.46\textwidth]{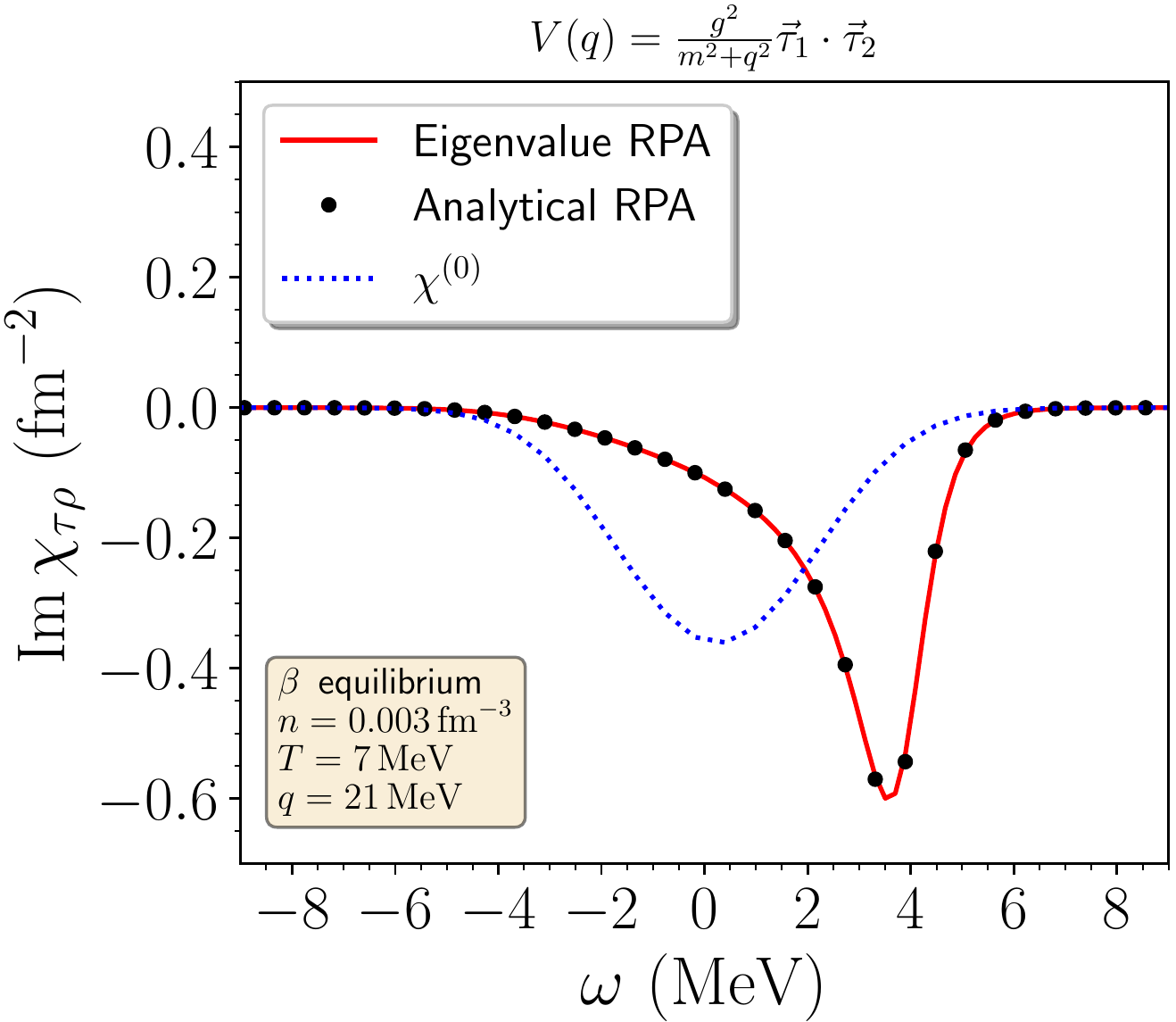}
\caption{Imaginary part of the charged-current density response function at $0^{\rm th}$ order (blue dots) and computed from the RPA resummation of the direct term analytically (black dots) and within the eigenvalue formalism described in Section \ref{sec:rpa} (red line). A smearing length $\epsilon = 70\times 10^{-4}\,{\rm fm}^{-1}$ in Eq.\ \eqref{smear} is employed.}
\label{chirhow}
\end{figure}

Our goal will be to extract the imaginary part of the vertex function, which is related to nuclear matter dynamical structure functions and neutrino scattering cross sections. In order for the imaginary part of Eq.\ \eqref{lmat} to be nonzero, the matrix $N^{-1}(E + \omega \mathbb{1})-{\cal V}$ must be singular, which occurs when $\omega$ takes on the values defined by
\begin{equation}
(NV-E) | \ell \rangle = \omega_\ell | \ell \rangle.
\end{equation}
In the vicinity of $\omega^\ell$, one can write
\begin{align}
& \hspace{-.2in} [ N^{-1} ( E + ( \omega + i \eta) \mathbb{1} ) - {\cal V} ]^{-1} 
\nonumber \\
& = \frac{1}{\langle \ell | N^{-1} | \ell \rangle} \left [ \frac{{\rm Pr}}{\omega - \omega_\ell} - i \pi \delta(\omega -\omega_\ell) \right ] | \ell \rangle \langle \ell |,
\end{align}
where ${\rm Pr}$ denotes the principal value. The imaginary part of the response function is then given by
\begin{equation}
{\rm Im}\, \chi_{\tau \rho}^{\rm RPA}(\vec q,\omega) = -i \pi \sum_\ell \frac{\langle B | \ell \rangle^2}{\langle \ell | N^{-1} | \ell \rangle}\delta(\omega - \omega_\ell).
\end{equation}
In terms of the discrete momentum-space mesh points and weights, we have
\begin{equation}
{\rm Im}\, \chi_{\tau \rho}^{\rm RPA}(\vec q,\omega) = -i \pi \sum_\ell \frac{ \left( \sum_i w_i | \ell \rangle_i \right )^2}{\sum_i (w_i | \ell \rangle_i)^2 (w_i N_i)^{-1}}\delta(\omega - \omega_\ell),
\label{rpamesh}
\end{equation}
where $| \ell \rangle_i$ denotes the $i^{\rm th}$ element of the eigenvector $| \ell \rangle$. In practice, the finite number of $\delta$ functions in Eq.\ \eqref{rpamesh} obtained by discretizing the integral in Eq.\ \eqref{inteq2} must be appropriately smeared to obtain a continuous response function. We employ the approximation
\begin{equation}
\delta_\epsilon(\omega) = \frac{1}{\sqrt{2 \pi \epsilon}}e^{-\omega^2/2 \epsilon}.
\label{smear}
\end{equation}
and find that it is always possible to choose a smearing length $\epsilon$ that leads to a converged result.

\begin{figure}[t]
\centering
\includegraphics[width=0.46\textwidth]{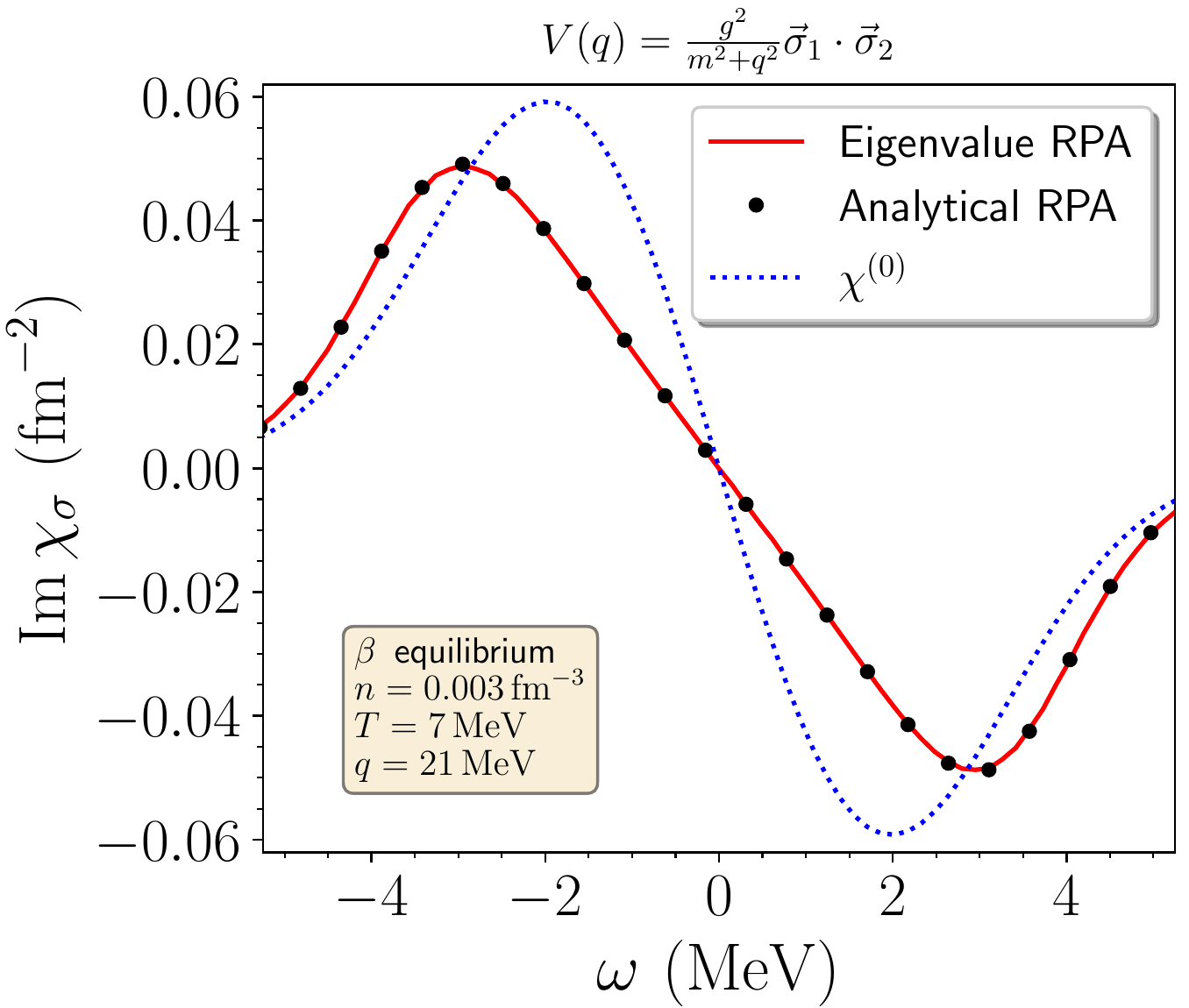}
\caption{Imaginary part of the neutral-current spin response function at $0^{\rm th}$ order (blue dots) and computed from the RPA resummation of the direct term analytically (black dots) and within the eigenvalue formalism described in Section \ref{sec:rpa} (red line). A smearing length $\epsilon = 70\times 10^{-4}\,{\rm fm}^{-1}$ in Eq.\ \eqref{smear} is employed.}
\label{chisigz}
\end{figure}

The above eigenvalue method is suitable to resum both the direct and exchange RPA bubble diagrams to all orders. To benchmark the method, we consider the simplified case of iterating just the direct part of the nuclear potential to all orders, which for a local potential $V(q)$ can be computed analytically through
\begin{equation}
\chi^{\rm RPA}(\vec q,\omega) = \frac{\chi^{(0)}(\vec q,\omega)}{1-V(q)\chi^{(0)}(\vec q,\omega)}.
\end{equation}
In Figure \ref{chirhow} we show the energy dependence of the charged-current density response function in beta-equilibrium nuclear matter at density $n=50 \times 10^{11}$\,g/cm$^3$ and temperature $T=7$\,MeV for a momentum transfer $q=21$\,MeV assuming a scalar-isovector interaction
\begin{equation}
V(q) = \frac{g^2}{m^2+q^2}\vec \tau_1 \cdot \vec \tau_2,
\end{equation}
where we take $g=5$ and $m=700$\,MeV. In Figure \ref{chirhow} we plot the response function in the approximation of noninteracting particles (blue dots), the exact analytical RPA result (black dots), and the numerical eigenvalue RPA result (red line) with a delta function smearing length of $\epsilon = 70\times 10^{-4}\,{\rm fm}^{-1}$. One sees excellent agreement between the exact and numerical RPA resummations.

The treatment of spin response functions proceeds similarly, except that we obtain a pair of coupled equations for the spin-up and spin-down response functions, which effectively doubles the dimensionality of the vectors and matrices in Eqs.\ \eqref{lmatrix} -- \eqref{vmatrix}. For the spin response functions, we also do not use the spin-averaging approximation employed to obtain Eq.\ \eqref{inteq2}. In Figure \ref{chisigz} we show the energy dependence of the neutral-current spin response function in beta-equilibrium nuclear matter at density $n=50 \times 10^{11}$\,g/cm$^3$ and temperature $T=7$\,MeV for a momentum transfer $q=21$\,MeV assuming an isoscalar spin-spin interaction
\begin{equation}
V(q) = \frac{g^2}{m^2+q^2}\vec \sigma_1 \cdot \vec \sigma_2,
\end{equation}
where we take $g=10$ and $m=700$\,MeV. In Figure \ref{chirhow} we plot the response function in the approximation of noninteracting particles (blue dots), the exact analytical RPA result (black dots), and the numerical eigenvalue RPA result (red line) with a delta function smearing length of $\epsilon = 70\times 10^{-4}\,{\rm fm}^{-1}$. Again we find excellent agreement between the exact and numerical RPA resummations.

\subsection{Dynamic structure functions}

Neutrino opacities are a key input to numerical simulations of core-collapse supernovae, proto-neutron star evolution, and neutron star mergers. Both neutral-current and charged-current weak reactions are important sources of neutrino opacity across a wide range of densities and temperatures. Matter effects on neutrino scattering and absorption cross sections on baryons are encoded in dynamical structure functions related to the imaginary part of nuclear response functions.


\subsubsection{Neutral-current neutrino scattering}

The double differential cross section for low-energy neutrinos to scatter in a non-relativistic gas of nucleons is given by 
\begin{eqnarray}
\label{diff}
&& \hspace{-.15in} \frac{1}{V}\frac{d^2 \sigma}{d \cos \theta\, d\omega} = \frac{G_F^2}{4\pi^2} (E_{\nu}-\omega)^2 \\ \nonumber 
&& \times \left [ c_V^2 (1 + \cos \theta ) S_{\rho}(q,\omega) 
+ c_A^2 (3 - \cos \theta) S_{\sigma}(q,\omega) \right ],
\end{eqnarray}
where $S_{\rho}$ is the neutral-current density structure function and $S_{\sigma}$ is the neutral-current spin structure function. The energy transfer is given by $\omega=E_\nu-E_\nu^\prime$ and the momentum transfer is given by $\vec q = \vec p_\nu - \vec p_\nu^{\, \prime}$ with magnitude $q=\sqrt{E_\nu^2 + {E_\nu^\prime}^2-2 E_\nu E_\nu^\prime \cos \theta}$.
The structure functions in Eq.\ \eqref{diff} are related to the imaginary parts of the associated response functions by
\begin{equation} \label{structure}
S (q,\omega) = - \frac{2\, {\rm Im} \chi(q,\omega)}{1 - e^{-\omega /T}}.
\end{equation}

\begin{figure}[t]
\includegraphics[width=0.95\linewidth]{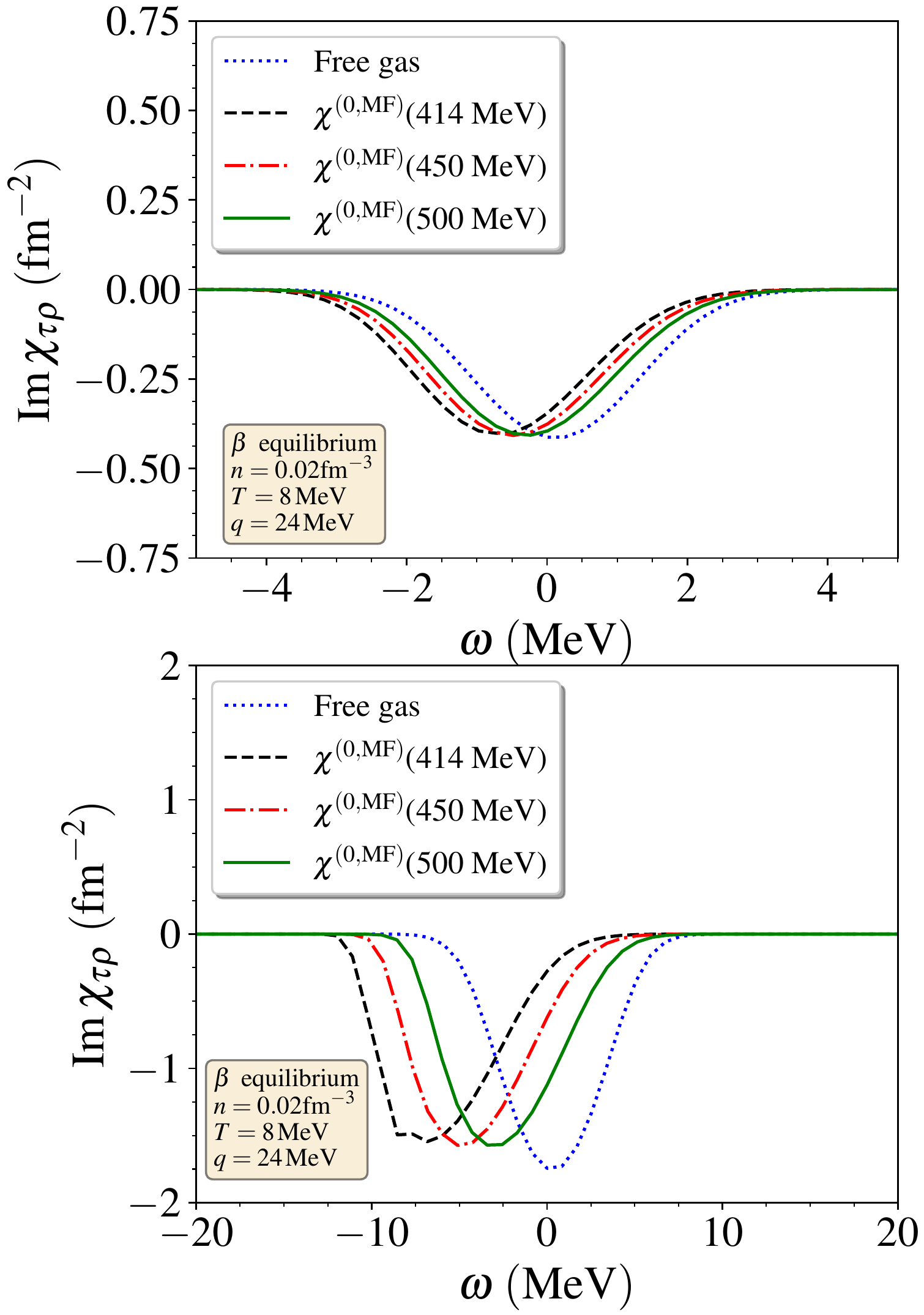}
\caption{Imaginary part of the charged-current density response function for a noninteracting gas (dotted blue line) and including the effects of nuclear mean fields in the Hartree-Fock approximation for different chiral nuclear potentials labeled by their momentum-space cutoff value: 414\,MeV, 450\,MeV, and 500\,MeV.}
\label{res:diffE}
\end{figure}

\begin{figure}[t]
\includegraphics[width=1.0\linewidth]{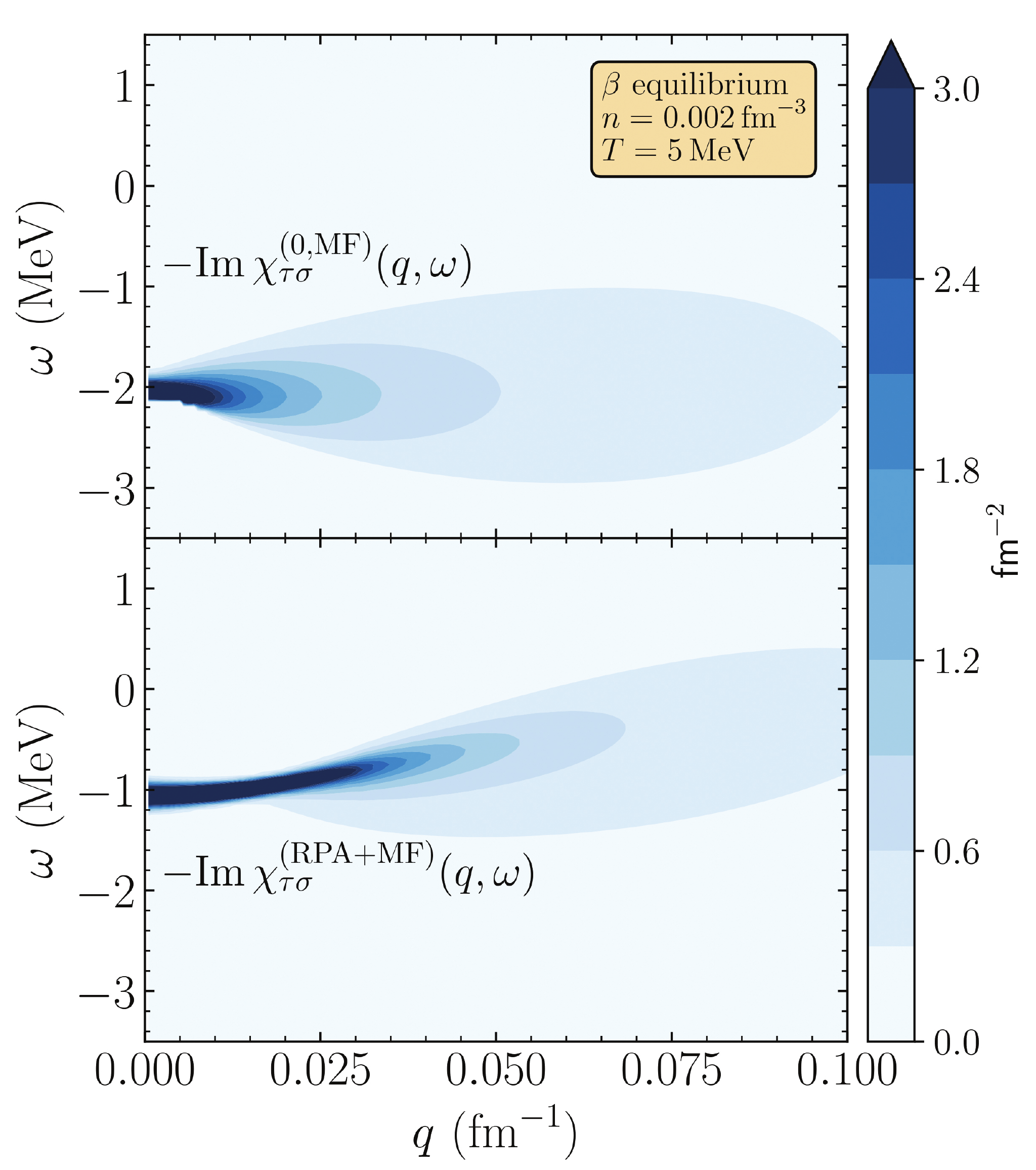}
\caption{Imaginary part of the charged-current spin response function of beta-equilibrium nuclear matter at a density $n=0.002$\,fm$^{-3}$ and temperature $T=5$\,MeV. (Top panel) spin response function including mean field ``MF'' corrections from the N3LO-414 chiral nuclear potential. (Bottom panel) spin response function including random phase approximation vertex corrections plus nuclear mean fields ``RPA+MF''.}
\label{ImChits002}
\end{figure}

\subsubsection{Charge-current neutrino absorption}

The double differential cross section for electron neutrino absorption  is given by \cite{roberts12,roberts17,iwamoto82}
\begin{eqnarray}
\label{scc}
    &&\hspace{-.2in} \frac{1}{V}\frac{d^2\sigma}{d\cos\theta dE_e} = \frac{G_F^2 \cos^2 \theta_c}{4\pi^2} p_e E_e (1-f_e(E_e))\\ \nonumber
    &&\times [(1+\cos\theta)S_{\tau \rho}(\omega,q) + g_A^2(3-\cos\theta)S_{\tau \sigma}(\omega,q)],
\end{eqnarray}
where $S_{\tau \rho}$ is the charged-current density structure function and $S_{\tau \sigma}$ is the charged-current spin structure function. The energy transfer is given by $\omega=E_\nu-E_e$ and the momentum transfer is given by $\vec q = \vec p_\nu - \vec p_e$ with magnitude $q=\sqrt{E_\nu^2 + E_e^2-2 E_\nu E_e \cos \theta}$. The structure functions in Eq.\ \eqref{scc} are related to the imaginary part of the associated response functions by
\begin{equation} \label{structure}
S_\tau(q,\omega) = -\frac{2\, {\rm Im} \chi (q,\omega)}{1 - e^{-(\omega + \mu_n - \mu_p) /T}},
\end{equation}
where the detailed balance factor depends explicitly on the proton and neutron chemical potentials $\mu_p$ and $\mu_n$.

\begin{figure}[t]
\includegraphics[width=1.0\linewidth]{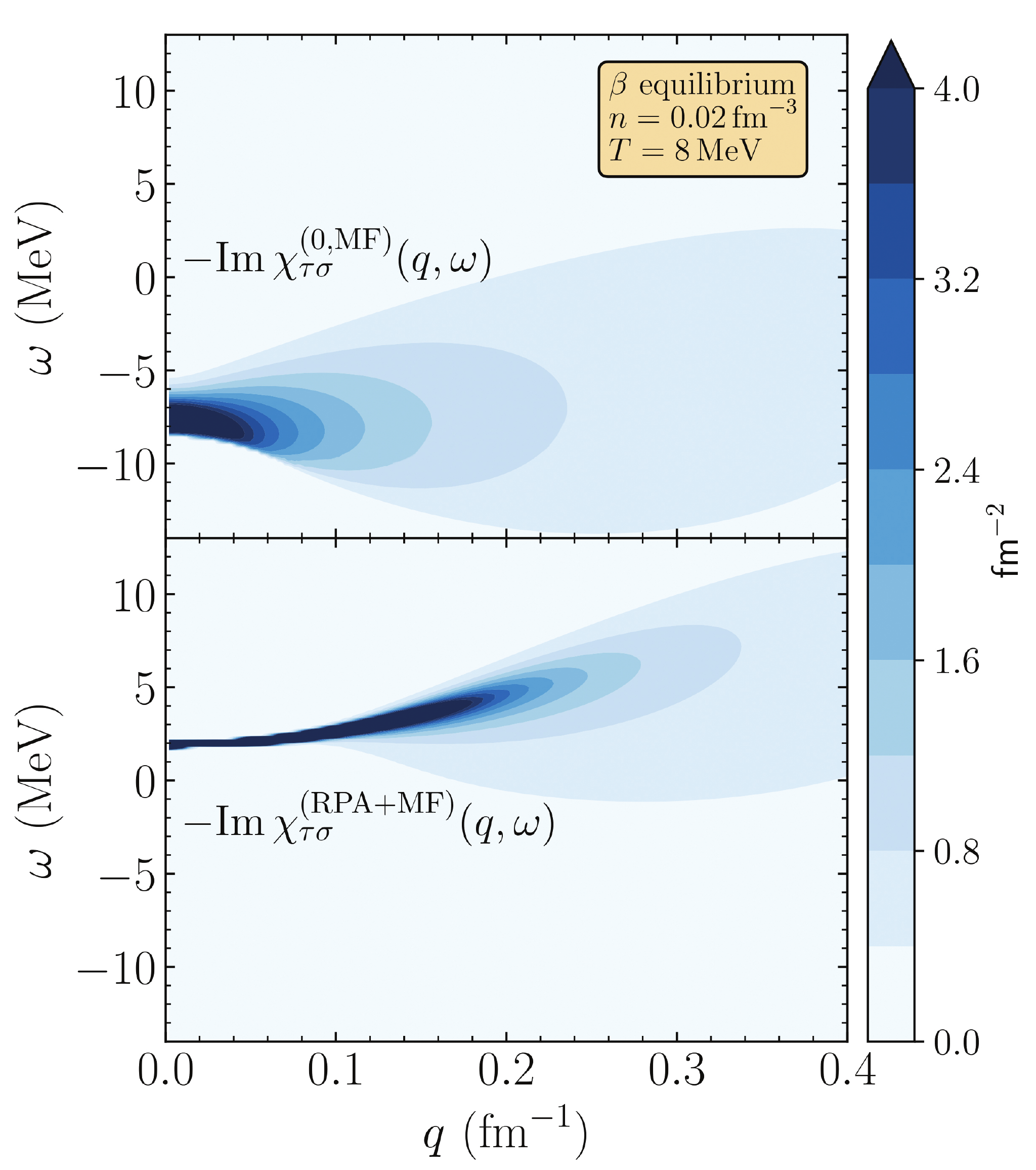}
\caption{Imaginary part of the charged-current spin response function of beta-equilibrium nuclear matter at a density $n=0.02$\,fm$^{-3}$ and temperature $T=8$\,MeV. (Top panel) spin response function including mean field ``MF'' corrections from the N3LO-414 chiral nuclear potential. (Bottom panel) spin response function including random phase approximation vertex corrections plus nuclear mean fields ``RPA+MF''.}
\label{ImChits02}
\end{figure}

\section{Results}
\label{res}

\subsection{Response functions in the mean field approximation}

\begin{figure*}[t]
\centering
\includegraphics[width=\textwidth]{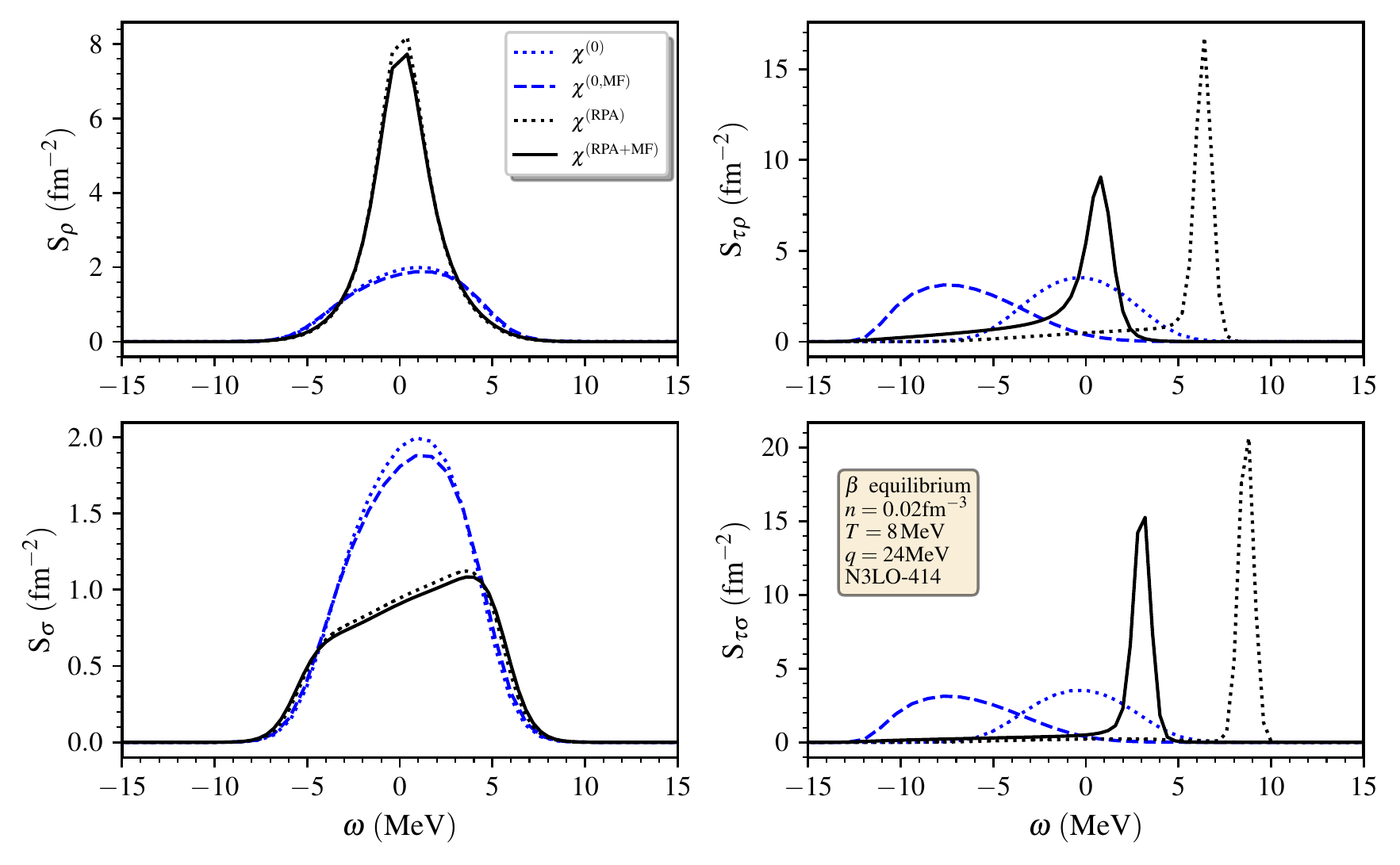}
\caption{Energy-dependent neutral-current (left) and charged-current (right) dynamic structure functions of beta-equilibrium nuclear matter at density $n = 0.02$\,fm$^{-3}$ and temperature $T=8$\,MeV for a momentum transfer of $q=24$\,MeV. We plot both the density (top) and spin (bottom) structure functions, including the effects of nucleon mean fields (dashed lines) and RPA correlations (black curves).}
\label{RPA_S}
\end{figure*}

In Figure \ref{res:diffE} we show mean field effects on the charged-current density response function of nuclear matter for different choices of the nuclear potential: N3LO-414, N3LO-450, and N3LO-500. In the top panel, we consider beta-equilibrium nuclear matter at density $n=0.002\,\text{fm}^{-3}$, temperature $T=5$\,MeV, and assuming a momentum transfer of $q=15$\,MeV. In the bottom panel, we consider beta-equilibrium nuclear matter at density $n=0.02\,\text{fm}^{-3}$, temperature $T=8$\,MeV, and assuming a momentum transfer of $q=24$\,MeV. We find that mean field effects are larger for smaller values of the momentum-space cutoff. A similar effect was observed in the context of the nuclear equation of state \cite{Holt} and the nuclear single-particle potential \cite{holt13,holt16}. Namely, low-cutoff potentials are more perturbative and therefore generate more attraction at first order in perturbation theory, even though the sum of first- and second-order perturbation theory contributions to the nuclear equation of state or single-particle potential are similar. As argued in Section \ref{meth}, the largest effect of mean field corrections in the imaginary part of the neutrino absorption charged-current response is to shift the strength by an amount $U_p-U_n$ in the energy transfer $\omega$. At the lower value of the density, $n=0.002\,\text{fm}^{-3}$, the mean field splitting is on the order of $1\,\text{MeV}$, while at the larger density $n=0.02\,\text{fm}^{-3}$, the mean field spliting is nearly $10\,\text{MeV}$ for the N3LO-414 chiral nucleon-nucleon potential. However, such mean field splittings are still about a factor of  2 less than those generated from typical phenomenological mean field models \cite{roberts12,rrapaj16}.


\subsection{Response functions in the random phase approximation}

The random phase approximation for the vertex function together with the Hartree-Fock approximation for the single-particle energies represents a conserving approximation that is guaranteed to preserve sum rules and the positivity of dynamical structure functions. In addition, it is able to capture the presence of collective oscillations such as the giant-dipole and Gamow-Teller resonances that are known to play a role in the response of nuclei \cite{reddy99}.   
In Figure \ref{ImChits002} we show a contour plot of the imaginary part of the neutrino-absorption charged-current spin response as a function of the energy $\omega$ and momentum $q$ transfer for beta-equilibrium matter with density $n=0.002\,\text{fm}^{-3}$ and temperature $T=5$\,MeV. The top panel shows the imaginary part of the response including mean field (MF) corrections alone, while the bottom panel shows the combined effect of RPA correlations and mean fields (RPA + MF). We see that the MF response exhibits a relatively broad distribution already at low momentum transfers that peaks at a nearly constant energy $\omega \simeq -2$\,MeV. In contrast, the RPA + MF response remains sharply peaked for longer, up to a momentum transfer $q \simeq 0.04\,\text{fm}^{-1} \simeq 8$\,MeV, and peaks at a value $\omega \simeq -1$\,MeV that increases slowly with $q$. The sharper structure of the RPA + MF response is indicative of a collective mode. This feature is even more evident in Figure \ref{ImChits002}, where we show the contour plot of the imaginary part of the neutrino-absorption charged-current spin response for beta-equilibrium matter with density $n=0.02\,\text{fm}^{-3}$ and temperature $T=8$\,MeV. Here the collective mode (now at positive energy) remains sharp up to a momentum transfer $q\simeq 0.2\,\text{fm}^{-1} \simeq 40$\,MeV. The shift in peak energy of the imaginary response from $-7.5$\,MeV in the mean field approximation to $2$\,MeV in the RPA + MF approximation will have an important effect on electron neutrino absorption in dilute beta-equilibrium nuclear matter. The shift will push the outgoing electron energy to smaller values where Pauli blocking acts to reduce the available phase space, thereby suppressing the absorption cross section and increasing the mean free path.

\begin{figure}[t]
\includegraphics[width=1.0\linewidth]{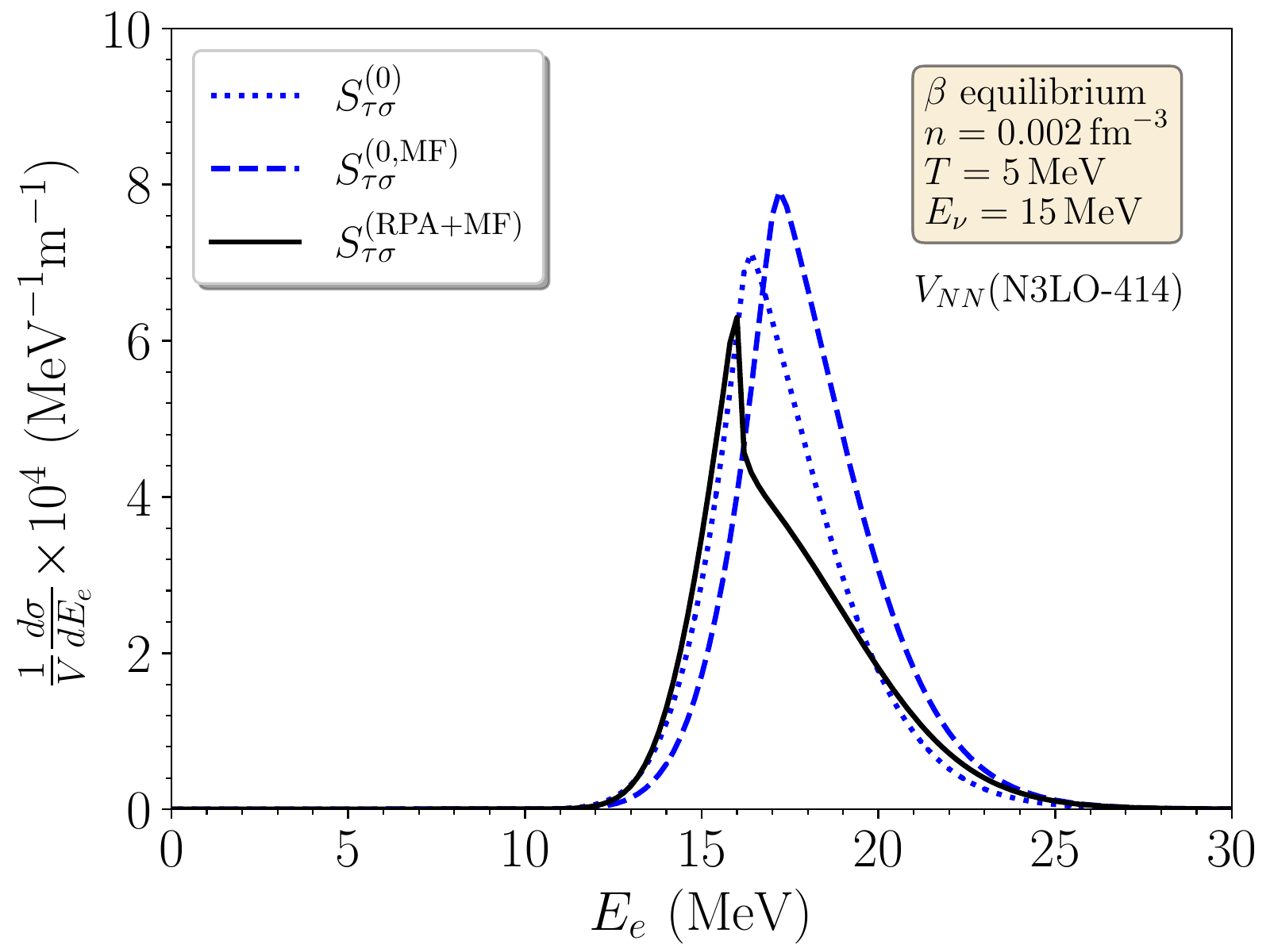}
\caption{Differential neutrino absorption cross section per unit volume as a function of the outgoing electron energy $E_e$ computed from the spin response function of beta-equilibrium nuclear matter at density $n=0.002$\,fm$^{-3}$ and temperature $T=5$\,MeV. The incoming neutrino energy is set to be $E_\nu = 3T = 15$\,MeV. Curves are shown for the case of noninteracting nucleons (blue dotted line), nuclear mean fields computed from the N3LO-414 chiral nuclear potential (blue dashed lines), and in the RPA + mean field approximation (black solid line) also employing the N3LO-414 potential.}
\label{scatt002}
\end{figure}

In the left panels of Figure \ref{RPA_S} we plot the neutral-current density (top) and spin (bottom) dynamic structure factors for beta equilibrium nuclear matter under the ambient conditions $T=8$\,MeV, $n=0.02$\,fm$^{-3}$, and momentum transfer of $q=24$\,MeV. For both the neutral-current density and spin response functions, the mean field corrections play only a very minor role. Mean fields change slightly the proton and neutron densities for beta equilibrium matter, and the effective mass has only a small effect on the Fermi distribution functions and energy denominators. The energy shifts in the single-particle potentials are absorbed into redefinitions of the proton and neutron chemical potentials for the Fermi distribution functions, and the mean field shifts cancel in the response function energy denominators. The inclusion of RPA correlations, however, is very important, enhancing the density structure function and suppressing the spin structure function, a feature already observed \cite{bedaque18} including first-order vertex corrections starting from a pseudopotential defined in terms of nucleon-nucleon scattering phase shifts. This behavior can be traced to the large neutron-neutron attraction in the $^1S_0$ partial wave.


\begin{figure}[t]
\includegraphics[width=1.0\linewidth]{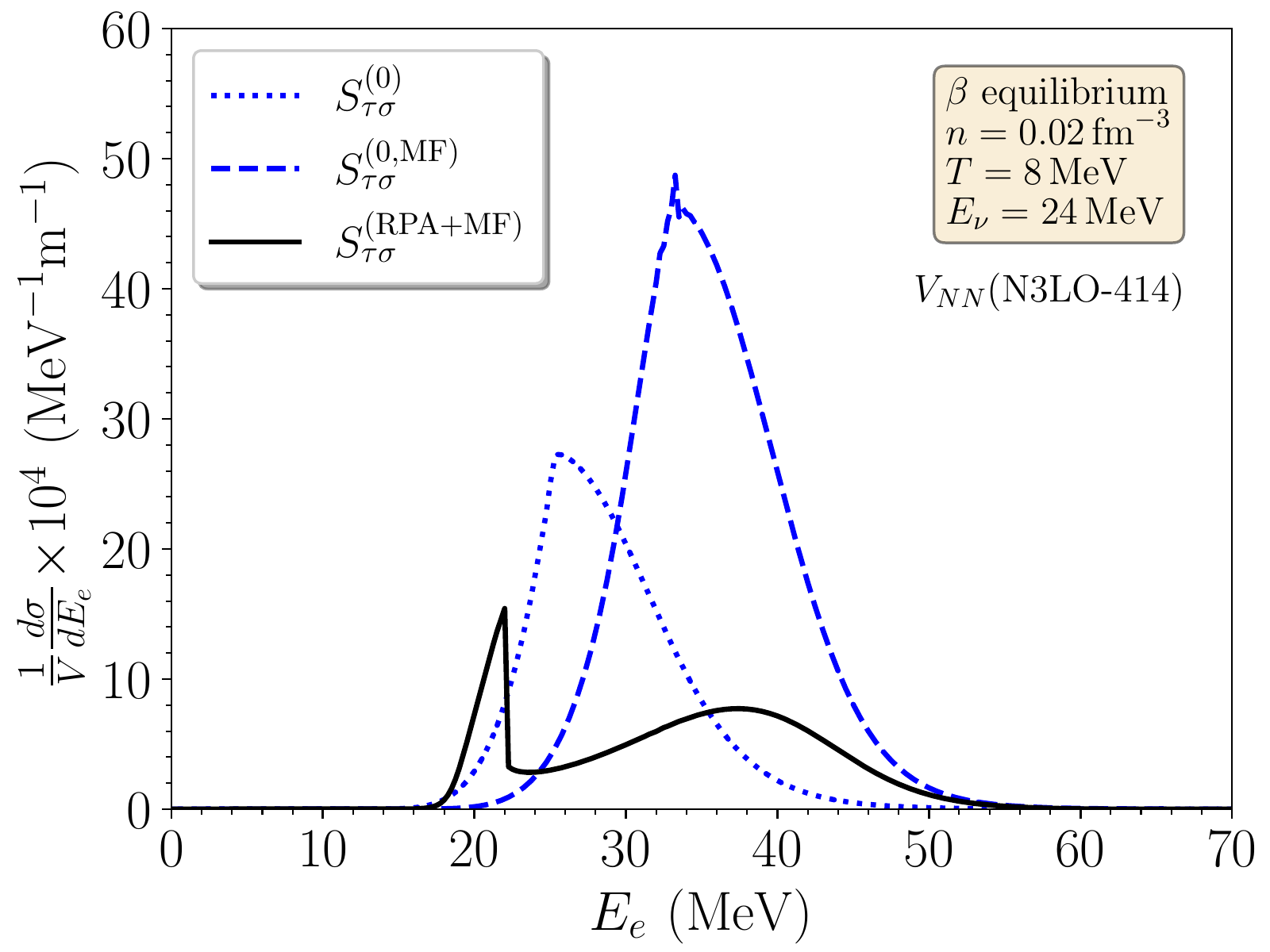}
\caption{Same as Figure \ref{scatt002} except for a density $n=0.02$\,fm$^{-3}$, temperature $T=8$\,MeV, and incoming neutrino energy $E_\nu = 24$\,MeV.}
\label{scatt02}
\end{figure}

In the right panels of Figure \ref{RPA_S} we plot the charged-current density (top) and spin (bottom) dynamic structure functions in beta-equilibrium nuclear matter under the ambient conditions $T=8$\,MeV, $n=0.02$\,fm$^{-3}$, and momentum transfer of $q=24$\,MeV. Whereas mean fields drive absorption strength to lower energy transfers, RPA correlations significantly shift strength to larger energy transfers. This is due to the existence of giant dipole and Gamow-Teller collective modes in the isovector-density and isovector-spin channels. The combined effect of nuclear mean fields and RPA correlations is a redistribution of strength to energies above that of the noninteracting charged-current response functions. We conclude that the inclusion of vertex corrections is crucial for an accurate description of nuclear matter response functions at and above neutrinosphere densities.



\subsection{Energy-dependent neutrino absorption cross section}

Integrating Eq.\ \eqref{scc} over the scattering angle $\theta$, we obtain the differential energy-dependent neutrino absorption cross section 
\begin{eqnarray}
\label{dsde}
    && \frac{1}{V}\frac{d\sigma}{dE_e} = \frac{G_F^2 \cos^2 \theta_c}{4\pi^2} p_e E_e (1-f_e(E_e))\\ \nonumber
    &&\times\int d\cos \theta [(1+\cos\theta)S_{\tau \rho}(\omega,q) + g_A^2(3-\cos\theta)S_{\tau \sigma}(\omega,q)],
\end{eqnarray}
where $S_{\tau \rho}$ and $S_{\tau \sigma}$ are the density and spin dynamic structure functions. In Figures \ref{scatt002} and \ref{scatt02} we plot the differential cross section for electron neutrino absorption, keeping only the spin dynamic structure function in Eq.\ \eqref{dsde} for beta-equilibrium matter at two densities $n = 0.002$\,fm$^{-3}$ and $n = 0.02$\,fm$^{-3}$. The incoming neutrino energy is $E_{\nu} = 3 T$. We show the cross section assuming noninteracting nucleons (blue dotted line), the cross section keeping only nuclear mean fields at the Hartree-Fock level (blue dashed line), and finally the cross section including RPA correlations and Hartree-Fock mean fields (black solid line). The nuclear force is taken to be the N3LO-414 chiral two-body interaction. One finds that the inclusion of nuclear mean fields enhances neutrino absorption since the response is shifted to lower energy transfers $\omega$ and, therefore, higher electron energies for which the Pauli suppression factor is reduced. However, RPA correlations provide a stronger shift in the response function toward higher energy transfers, leading to a reduced absorption cross section whose peak in electron energy can be shifted below that of the non-interacting Fermi gas. 


\subsection{Mean free path}
\begin{figure}[t]
\includegraphics[width=\linewidth]{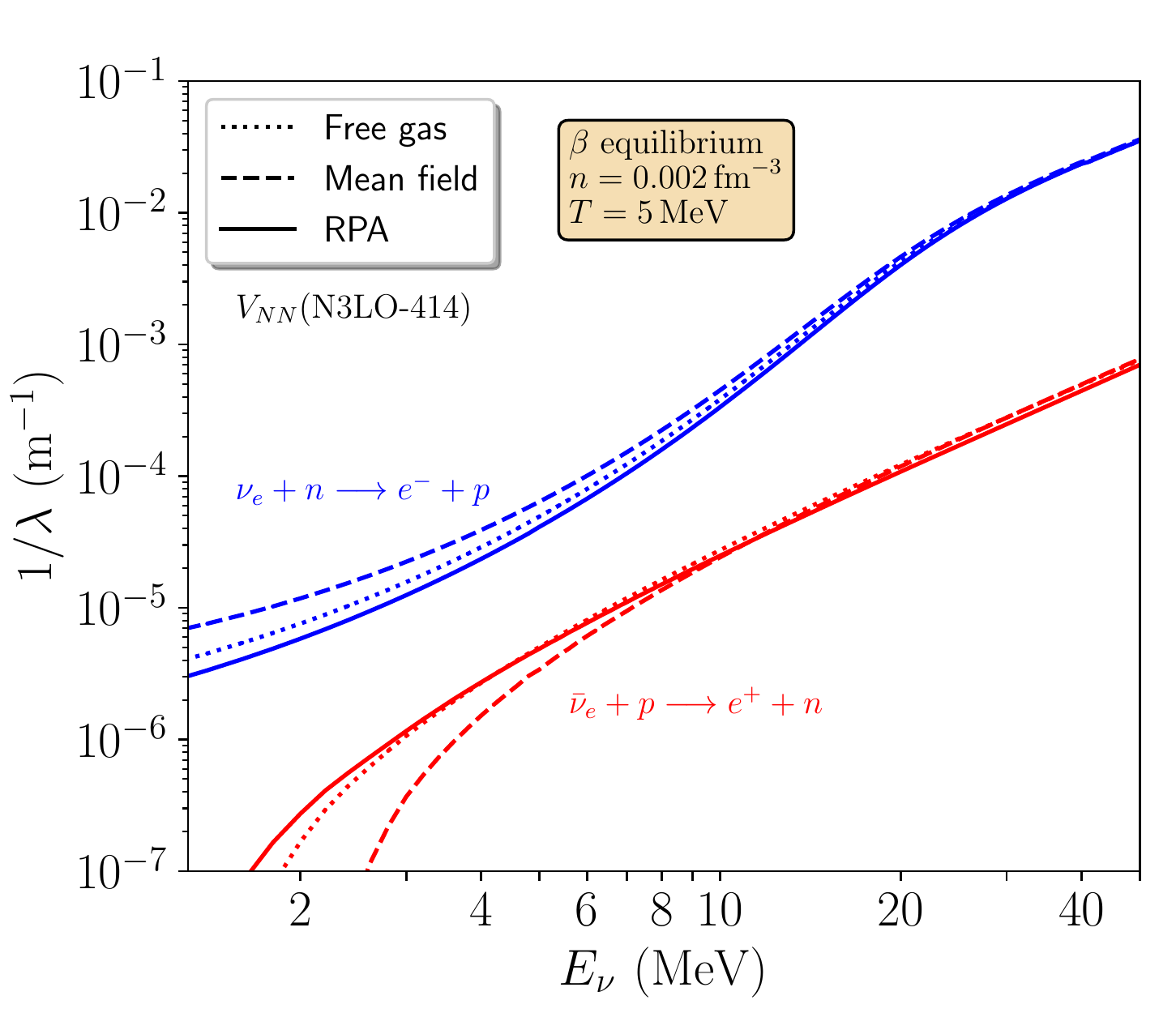}
\caption{Inverse neutrino and antineutrino absorption mean free paths as a function of energy $E_\nu$ in beta-equilibrium nuclear matter at density $n=0.002$\,fm$^{-3}$ and temperature $T=5$\,MeV. Shown are results for (i) noninteracting nucleons (dotted lines), (ii) mean field corrections (dashed lines), and (ii) RPA + mean field corrections (solid lines) all calculated using the N3LO-414 chiral nucleon-nucleon potential.}
\label{invlamb002}
\end{figure}
\begin{figure}[t]
\includegraphics[width=\linewidth]{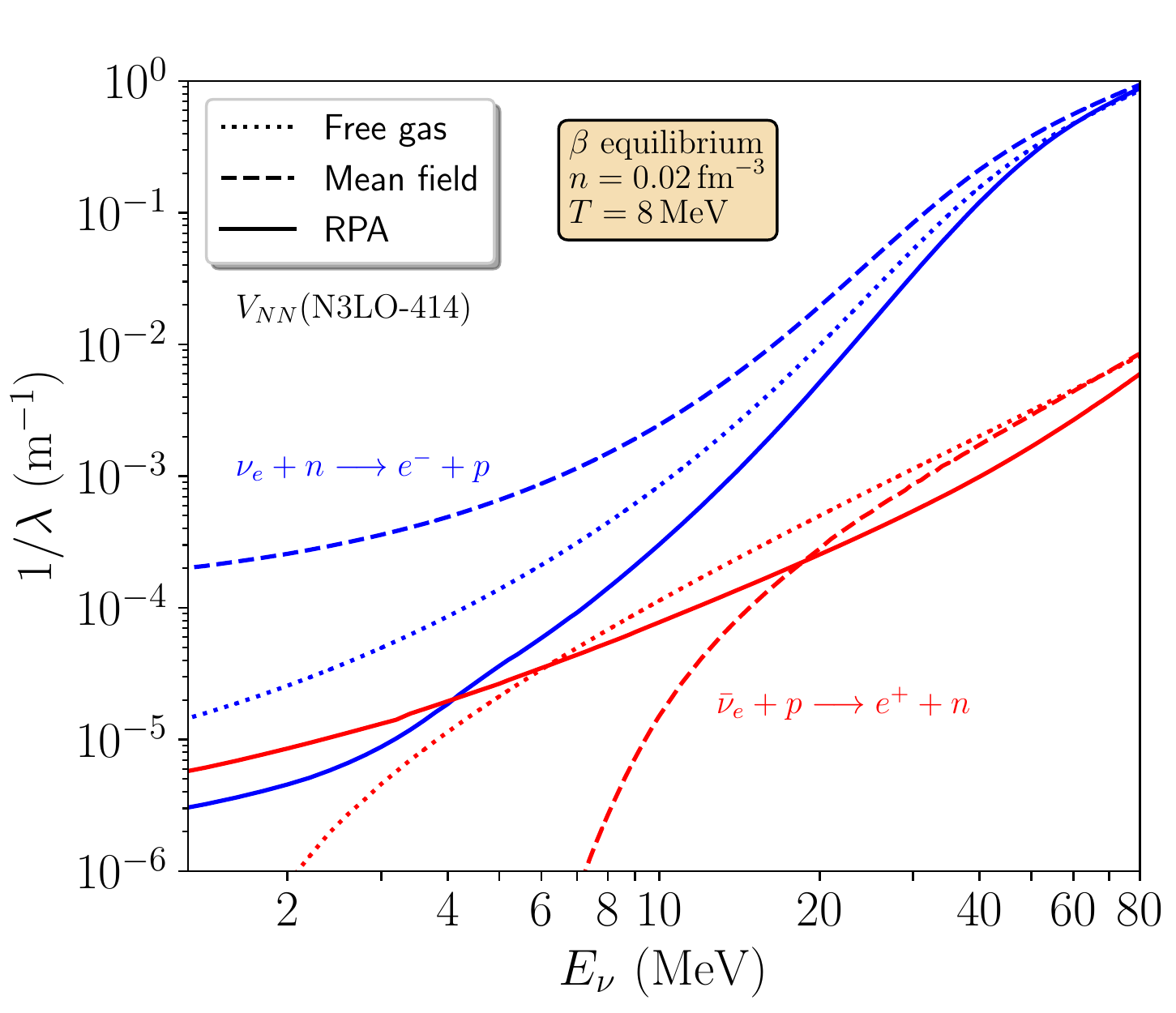}
\caption{Inverse neutrino and antineutrino absorption mean free paths as a function of energy $E_\nu$ in beta-equilibrium nuclear matter at density $n=0.02$\,fm$^{-3}$ and temperature $T=8$\,MeV. Shown are results for (i) noninteracting nucleons (dotted lines), (ii) mean field corrections (dashed lines), and (ii) RPA + mean field corrections (solid lines) all calculated using the N3LO-414 chiral nucleon-nucleon potential.}
\label{invlamb02}
\end{figure}
 We now present and discuss results for the mean free path of electron and anti-electron neutrinos in the vicinity of the neutrinosphere due to charged current interactions. Differences between these mean free paths directly impact several key observable aspects of supernovae and neutron star mergers including dynamics, nucleosynthesis, and neutrino oscillations. 
 
 The inverse of the electron and anti-electron neutrino  mean free path due to their charged current interactions is obtained by integrating the differential absorption cross section per unit volume over the final-state lepton energy, e.g.,
\begin{equation}
    \frac{1}{\lambda}= \int  \frac{1}{V}\frac{d\sigma}{dE_e} dE_e\,. 
\end{equation}

In Figures \ref{invlamb002} and \ref{invlamb02} we plot the inverse mean free paths of electron neutrino and antineutrino absorption as a function of the incident energy for two sets of ambient conditions $(n,T)=(0.002\,\text{fm}^{-3},5\,\text{MeV})$ and $(0.02\,\text{fm}^{-3},8\,\text{MeV})$. The dynamic structure functions are computed from the associated charged-current spin response functions in three approximations. First, the inverse neutrino mean free paths neglecting interactions between nucleons is shown by the dotted curves. Second, we show as the dashed lines the effect of introducing proton and neutron mean fields in the Hartree-Fock approximation employing the N3LO-414 chiral nucleon-nucleon interaction. Finally, the solid curves show the combined effects of nucleon mean fields and vertex corrections obtained in the random phase approximation. We find that for the N3LO-414 potential, the mean-field effects significantly enhance the electron neutrino absorption cross-section across all energies considered in agreement with earlier studies. However, in contrast, RPA correlations redistribute strength to the vicinity of the positive-energy collective mode. This significantly reduces the outgoing electron energy into a region where Pauli blocking suppresses the reaction. This redistribution of strength due to a broad collective mode shifts the response to higher energy and undoes the enhancement of the inverse mean free path due to mean-field effects. Remarkably, correlations suppress the electron neutrino absorption cross-sections over the entire energy range and are especially large for low-energy neutrinos. 

\begin{figure}[t]
\includegraphics[width=1.0\linewidth]{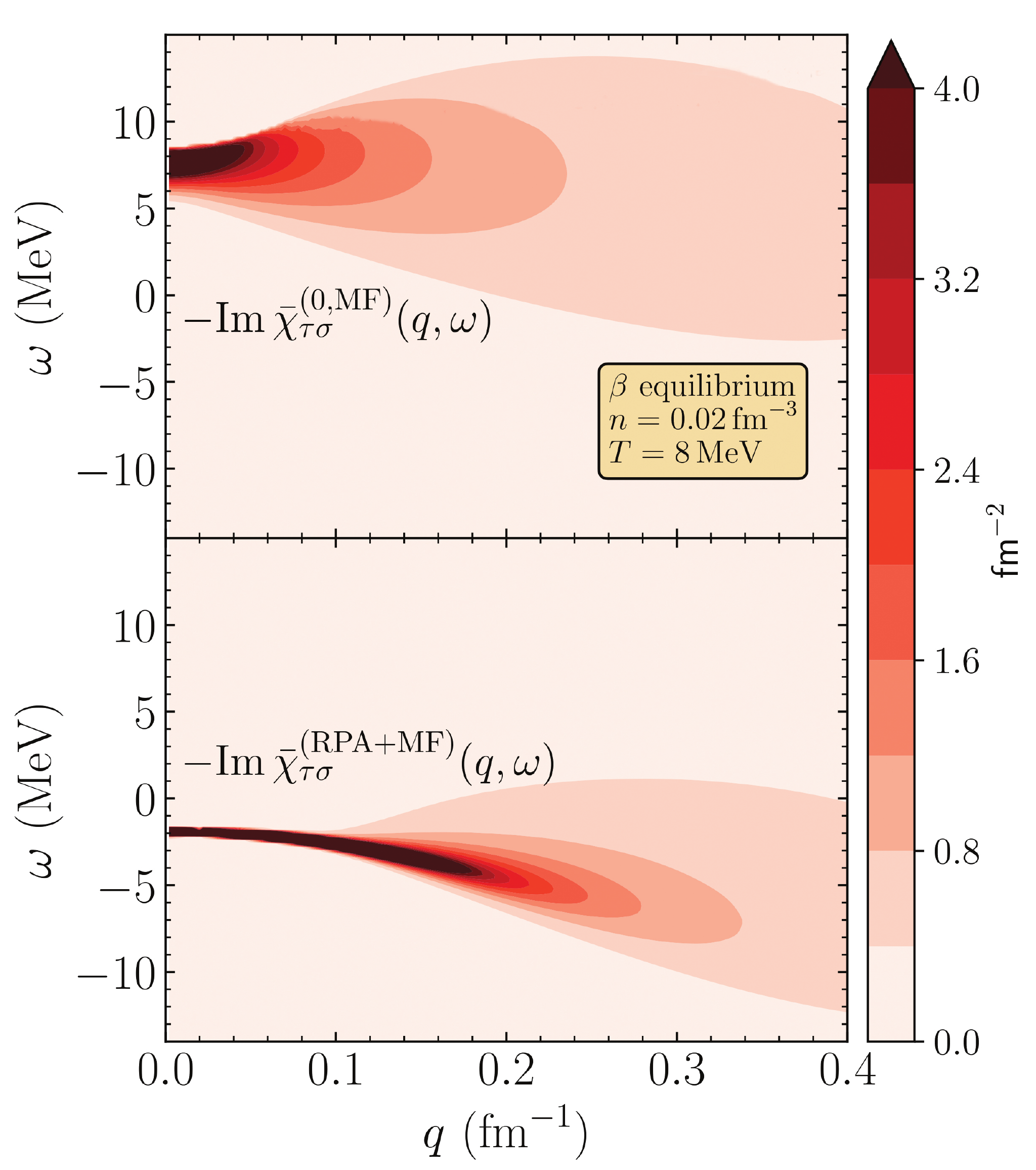}
\caption{Imaginary part of the antineutrino absorption charged-current spin response function $\bar 
\chi_{\tau \sigma}$ of beta-equilibrium nuclear matter at a density $n=0.02$\,fm$^{-3}$ and temperature $T=8$\,MeV. (Top panel) spin response function including mean field ``MF'' corrections from the N3LO-414 chiral nuclear potential. (Bottom panel) spin response function including random phase approximation vertex corrections plus nuclear mean fields ``RPA+MF''.}
\label{ANuImChits02}
\end{figure}

In Figure \ref{ANuImChits02} we show the imaginary part of the spin response function for antineutrino absorption in beta-equilibrium nuclear matter at density $n=0.02\,\text{fm}^{-3}$ and temperature $T = 8\,\text{MeV}$. For electron antineutrinos, nuclear mean fields shift the response to higher energies. The absorption cross-section is therefore greatly reduced because the threshold energy to convert protons into neutrons is increased to such an extent that there is little phase space available for the reaction. As observed in Figure \ref{invlamb02}, the corresponding mean free path increases dramatically at low energies in agreement with earlier work. However, the presence of the negative-energy collective mode due to RPA correlations, shown in the lower panel of Figure \ref{ANuImChits02}, lowers the energy required for the process. It provides a reaction pathway even at low antineutrino energies, thereby increasing the cross-section. The reaction at low energies can be viewed as a process involving the absorption of a positively charged collective mode by the anti-electron neutrino to produce a positron in the final state. At higher antineutrino energies, the strong coupling to the negative-energy collective mode weakens the absorption cross section through the detailed balance factor $\left ( 1 - e^{-(\omega + \mu_p - \mu_n) /T} \right )^{-1}$. The response function is also narrowly peaked in this region, which limits the available phase space for final-state positron energies. Overall, at high energies the antineutrino absorption total cross section is reduced relative to both the noninteracting and mean field approximations as seen in Figure \ref{invlamb02}.


\section{Conclusion}
\label{con}
We have developed the framework to calculate the neutrino scattering and absorption rates in a warm neutron-rich matter that consistently includes mean-field effects and correlations through the Random Phase Approximation (RPA). We employ nuclear interactions derived from chiral effective field theory and, for the first time, include direct and exchange contributions to the mean field energies and RPA vertex functions.  The combination of Hartree-Fock self energies and RPA vertex corrections to the response function constitutes a ``conserving approximation'' that guarantees thermodynamic consistency and the positivity of dynamic structure functions. The integral equation for the RPA vertex function was discretized, leading to a matrix eigenvalue problem that could be solved through standard diagonalization. 

We find that including RPA correlations produces a broad collective mode that shifts the strength of the charged-current neutrino response to higher energy transfer. For a fixed incident neutrino energy, this lowers the outgoing electron energy into a region of Pauli-blocked suppression, thereby reducing the total cross-section. In contrast, electron-antineutrino absorption is enhanced at low energies because the same broad collective mode lowers the threshold energy needed to convert protons into neutrons. For higher antineutrino energy, this enhancement is absent because of kinematic constraints.

Our consistent treatment of nuclear mean fields and RPA correlations has important implications for charged current reactions near the supernova and neutron star merger neutrinospheres. 
When only mean-field effects are included, we confirm the results of previous studies that found a large enhancement of electron-neutrino absorption cross-section and a reduction in antineutrino absorption cross-section.  However, correlations included through RPA qualitatively change the picture. As discussed in the previous section, the absorption cross-sections for electron neutrinos are reduced over the entire range of relevant energies. This would imply an increased luminosity and average energy of electron neutrinos emitted in supernovae and mergers. For anti-electron neutrinos, the absorption cross-section is enhanced at low energy and suppressed at high energy. This implies that the lepton number flux carried by neutrinos can be strongly energy-dependent when RPA correlations are included. Since neutrino oscillations and nucleosynthesis are especially sensitive to the neutrino lepton number flux and its energy dependence \cite{Qian1996, Arcones2013}, our findings will likely impact both. 

Our finding that vertex corrections suppress differences between $\nu_e$ and $\bar{\nu}_e$ absorption rates in the vicinity of the neutrinosphere implies that the emerging spectra of $\nu_e$ and $\bar{\nu}_e$ will be more similar than previously expected \cite{martinez-pinedo12,roberts12}. This will likely inhibit the production of neutron-rich r-process nuclei in the neutrino driven ejecta from core-collapse supernovae and dynamical ejecta in neutron star mergers (for recent reviews see \cite{Kajino:2019abv,Cowan:2019pkx}). However, collective neutrino flavor oscillations, which are also sensitive to the energy dependence of the lepton number flux, impact the final spectra at the nucleosynthesis site (see reviews~\cite{Chakraborty2016,Tamborra2021} and references therein). To gauge if corrections to mean free paths calculated in this work can alter  nucleosynthesis it is critical to include their effect on flavor transformation by modifying the two-point Feynman diagrams for electron neutrino absorption and emission processes in Quantum Kinetic Equation (QKE) treatments~\cite{Vlasenko2014,Richers2019}, and collisional instabilities~\cite{Johns2022,Johns2023}. Further, larger mean free paths for both electron and anti-electron neutrinos at higher energy imply an increase in the total luminosity and average energy.  This will likely increase the net neutrino energy deposition behind the shock in core-collapse supernova and aid the explosion. The results of this work can be tabulated and included in simulations of core-collapse supernovae, neutron star mergers, and neutron star cooling.              

Finally, we mention the limitations of our study and identify directions for future work. First, although our analysis provides a consistent treatment of excitations above the Hartree-Fock ground state, it neglects two-body currents and correlations beyond one-particle-one-hole RPA. At the low density, we expect both two-body currents and two-particle-two-hole excitations to be small because they appear at higher order in the density expansion, but more work is needed to assess their importance at densities of relevance to the neutrino sphere. It is well-known that the interplay between short-range correlations and two-body currents, especially in the axial vector channel, plays an important role in nuclear weak interactions. In addition, error estimates for neutrino interaction rates require systematic order-by-order calculations. These  quantitive issues warrant further work before one can draw definite conclusions about $\nu_e$ and $\bar{\nu}_e$ charged current reactions in the neutrino sphere and their emergent spectra.

\section{Acknowledgement}
The work of E.\ Shin and J.\ W.\ Holt is supported by the National Science Foundation under Grant Nos.\ PHY1652199 and PHY2209318. Portions of the research were conducted with the advanced computing resources provided by Texas A\&M High Performance Research Computing. The work of S. R. was supported by the U.S. DOE under Grant No. DE-FG02-00ER41132 and by the National Science Foundation's Physics Frontier Center: The Network for Neutrinos, Nuclear Astrophysics, and Symmetries.

\bibliographystyle{apsrev4-1}

\begin{thebibliography}{52}%
\makeatletter
\providecommand \@ifxundefined [1]{%
 \@ifx{#1\undefined}
}%
\providecommand \@ifnum [1]{%
 \ifnum #1\expandafter \@firstoftwo
 \else \expandafter \@secondoftwo
 \fi
}%
\providecommand \@ifx [1]{%
 \ifx #1\expandafter \@firstoftwo
 \else \expandafter \@secondoftwo
 \fi
}%
\providecommand \natexlab [1]{#1}%
\providecommand \enquote  [1]{``#1''}%
\providecommand \bibnamefont  [1]{#1}%
\providecommand \bibfnamefont [1]{#1}%
\providecommand \citenamefont [1]{#1}%
\providecommand \href@noop [0]{\@secondoftwo}%
\providecommand \href [0]{\begingroup \@sanitize@url \@href}%
\providecommand \@href[1]{\@@startlink{#1}\@@href}%
\providecommand \@@href[1]{\endgroup#1\@@endlink}%
\providecommand \@sanitize@url [0]{\catcode `\\12\catcode `\$12\catcode
  `\&12\catcode `\#12\catcode `\^12\catcode `\_12\catcode `\%12\relax}%
\providecommand \@@startlink[1]{}%
\providecommand \@@endlink[0]{}%
\providecommand \url  [0]{\begingroup\@sanitize@url \@url }%
\providecommand \@url [1]{\endgroup\@href {#1}{\urlprefix }}%
\providecommand \urlprefix  [0]{URL }%
\providecommand \Eprint [0]{\href }%
\providecommand \doibase [0]{http://dx.doi.org/}%
\providecommand \selectlanguage [0]{\@gobble}%
\providecommand \bibinfo  [0]{\@secondoftwo}%
\providecommand \bibfield  [0]{\@secondoftwo}%
\providecommand \translation [1]{[#1]}%
\providecommand \BibitemOpen [0]{}%
\providecommand \bibitemStop [0]{}%
\providecommand \bibitemNoStop [0]{.\EOS\space}%
\providecommand \EOS [0]{\spacefactor3000\relax}%
\providecommand \BibitemShut  [1]{\csname bibitem#1\endcsname}%
\let\auto@bib@innerbib\@empty
\bibitem [{\citenamefont {Burrows}\ and\ \citenamefont
  {Sawyer}(1999)}]{burrows99}%
  \BibitemOpen
  \bibfield  {author} {\bibinfo {author} {\bibfnamefont {A.}~\bibnamefont
  {Burrows}}\ and\ \bibinfo {author} {\bibfnamefont {R.~F.}\ \bibnamefont
  {Sawyer}},\ }\href {\doibase 10.1103/PhysRevC.59.510} {\bibfield  {journal}
  {\bibinfo  {journal} {Phys. Rev. C}\ }\textbf {\bibinfo {volume} {59}},\
  \bibinfo {pages} {510} (\bibinfo {year} {1999})}\BibitemShut {NoStop}%
\bibitem [{\citenamefont {O'Connor}(2015)}]{oconnor15}%
  \BibitemOpen
  \bibfield  {author} {\bibinfo {author} {\bibfnamefont {E.}~\bibnamefont
  {O'Connor}},\ }\href {\doibase 10.1088/0067-0049/219/2/24} {\bibfield
  {journal} {\bibinfo  {journal} {Astrophys. J. Suppl.}\ }\textbf {\bibinfo
  {volume} {219}},\ \bibinfo {pages} {24} (\bibinfo {year} {2015})}\BibitemShut
  {NoStop}%
\bibitem [{\citenamefont {Melson}\ \emph {et~al.}(2015)\citenamefont {Melson},
  \citenamefont {Janka}, \citenamefont {Bollig}, \citenamefont {Hanke},
  \citenamefont {Marek},\ and\ \citenamefont {M\"uller}}]{melson15}%
  \BibitemOpen
  \bibfield  {author} {\bibinfo {author} {\bibfnamefont {T.}~\bibnamefont
  {Melson}}, \bibinfo {author} {\bibfnamefont {H.-T.}\ \bibnamefont {Janka}},
  \bibinfo {author} {\bibfnamefont {R.}~\bibnamefont {Bollig}}, \bibinfo
  {author} {\bibfnamefont {F.}~\bibnamefont {Hanke}}, \bibinfo {author}
  {\bibfnamefont {A.}~\bibnamefont {Marek}}, \ and\ \bibinfo {author}
  {\bibfnamefont {B.}~\bibnamefont {M\"uller}},\ }\href {\doibase
  {10.1088/2041-8205/808/2/L42}} {\bibfield  {journal} {\bibinfo  {journal}
  {Astrophys. J. Lett.}\ }\textbf {\bibinfo {volume} {808}},\ \bibinfo {pages}
  {L42} (\bibinfo {year} {2015})}\BibitemShut {NoStop}%
\bibitem [{\citenamefont {Roberts}\ \emph {et~al.}(2016)\citenamefont
  {Roberts}, \citenamefont {Ott}, \citenamefont {Haas}, \citenamefont
  {O'Connor}, \citenamefont {Diener},\ and\ \citenamefont
  {Schnetter}}]{roberts16}%
  \BibitemOpen
  \bibfield  {author} {\bibinfo {author} {\bibfnamefont {L.~F.}\ \bibnamefont
  {Roberts}}, \bibinfo {author} {\bibfnamefont {C.~D.}\ \bibnamefont {Ott}},
  \bibinfo {author} {\bibfnamefont {R.}~\bibnamefont {Haas}}, \bibinfo {author}
  {\bibfnamefont {E.~P.}\ \bibnamefont {O'Connor}}, \bibinfo {author}
  {\bibfnamefont {P.}~\bibnamefont {Diener}}, \ and\ \bibinfo {author}
  {\bibfnamefont {E.}~\bibnamefont {Schnetter}},\ }\href {\doibase
  10.3847/0004-637X/831/1/98} {\bibfield  {journal} {\bibinfo  {journal}
  {Astrophys. J.}\ }\textbf {\bibinfo {volume} {831}},\ \bibinfo {pages} {98}
  (\bibinfo {year} {2016})}\BibitemShut {NoStop}%
\bibitem [{\citenamefont {Pons}\ \emph {et~al.}(1999)\citenamefont {Pons},
  \citenamefont {Reddy}, \citenamefont {Prakash}, \citenamefont {Lattimer},\
  and\ \citenamefont {Miralles}}]{pons98}%
  \BibitemOpen
  \bibfield  {author} {\bibinfo {author} {\bibfnamefont {J.~A.}\ \bibnamefont
  {Pons}}, \bibinfo {author} {\bibfnamefont {S.}~\bibnamefont {Reddy}},
  \bibinfo {author} {\bibfnamefont {M.}~\bibnamefont {Prakash}}, \bibinfo
  {author} {\bibfnamefont {J.~M.}\ \bibnamefont {Lattimer}}, \ and\ \bibinfo
  {author} {\bibfnamefont {J.~A.}\ \bibnamefont {Miralles}},\ }\href {\doibase
  10.1086/306889} {\bibfield  {journal} {\bibinfo  {journal} {Astrophys. J.}\
  }\textbf {\bibinfo {volume} {513}},\ \bibinfo {pages} {780} (\bibinfo {year}
  {1999})}\BibitemShut {NoStop}%
\bibitem [{\citenamefont {Roberts}\ and\ \citenamefont
  {Reddy}(2017)}]{roberts17}%
  \BibitemOpen
  \bibfield  {author} {\bibinfo {author} {\bibfnamefont {L.~F.}\ \bibnamefont
  {Roberts}}\ and\ \bibinfo {author} {\bibfnamefont {S.}~\bibnamefont
  {Reddy}},\ }\href {\doibase 10.1103/PhysRevC.95.045807} {\bibfield  {journal}
  {\bibinfo  {journal} {Phys. Rev. C}\ }\textbf {\bibinfo {volume} {95}},\
  \bibinfo {pages} {045807} (\bibinfo {year} {2017})}\BibitemShut {NoStop}%
\bibitem [{\citenamefont {Sekiguchi}(2010)}]{sekiguchi10}%
  \BibitemOpen
  \bibfield  {author} {\bibinfo {author} {\bibfnamefont {Y.}~\bibnamefont
  {Sekiguchi}},\ }\href {\doibase 10.1088/0264-9381/27/11/114107} {\bibfield
  {journal} {\bibinfo  {journal} {Class. Quant. Grav.}\ }\textbf {\bibinfo
  {volume} {27}},\ \bibinfo {pages} {114107} (\bibinfo {year}
  {2010})}\BibitemShut {NoStop}%
\bibitem [{\citenamefont {{Wanajo}}\ \emph {et~al.}(2014)\citenamefont
  {{Wanajo}}, \citenamefont {{Sekiguchi}}, \citenamefont {{Nishimura}},
  \citenamefont {{Kiuchi}}, \citenamefont {{Kyutoku}},\ and\ \citenamefont
  {{Shibata}}}]{wanajo14}%
  \BibitemOpen
  \bibfield  {author} {\bibinfo {author} {\bibfnamefont {S.}~\bibnamefont
  {{Wanajo}}}, \bibinfo {author} {\bibfnamefont {Y.}~\bibnamefont
  {{Sekiguchi}}}, \bibinfo {author} {\bibfnamefont {N.}~\bibnamefont
  {{Nishimura}}}, \bibinfo {author} {\bibfnamefont {K.}~\bibnamefont
  {{Kiuchi}}}, \bibinfo {author} {\bibfnamefont {K.}~\bibnamefont {{Kyutoku}}},
  \ and\ \bibinfo {author} {\bibfnamefont {M.}~\bibnamefont {{Shibata}}},\
  }\href {\doibase 10.1088/2041-8205/789/2/L39} {\bibfield  {journal} {\bibinfo
   {journal} {Astrophys. J. Lett.}\ }\textbf {\bibinfo {volume} {789}},\
  \bibinfo {pages} {L39} (\bibinfo {year} {2014})}\BibitemShut {NoStop}%
\bibitem [{\citenamefont {Endrizzi}\ \emph {et~al.}(2020)\citenamefont
  {Endrizzi}, \citenamefont {Perego}, \citenamefont {Fabbri}, \citenamefont
  {Branca}, \citenamefont {Radice}, \citenamefont {Bernuzzi}, \citenamefont
  {Giacomazzo}, \citenamefont {Pederiva},\ and\ \citenamefont
  {Lovato}}]{endrizzi20}%
  \BibitemOpen
  \bibfield  {author} {\bibinfo {author} {\bibfnamefont {A.}~\bibnamefont
  {Endrizzi}}, \bibinfo {author} {\bibfnamefont {A.}~\bibnamefont {Perego}},
  \bibinfo {author} {\bibfnamefont {F.~M.}\ \bibnamefont {Fabbri}}, \bibinfo
  {author} {\bibfnamefont {L.}~\bibnamefont {Branca}}, \bibinfo {author}
  {\bibfnamefont {D.}~\bibnamefont {Radice}}, \bibinfo {author} {\bibfnamefont
  {S.}~\bibnamefont {Bernuzzi}}, \bibinfo {author} {\bibfnamefont
  {B.}~\bibnamefont {Giacomazzo}}, \bibinfo {author} {\bibfnamefont
  {F.}~\bibnamefont {Pederiva}}, \ and\ \bibinfo {author} {\bibfnamefont
  {A.}~\bibnamefont {Lovato}},\ }\href {\doibase
  10.1140/epja/s10050-019-00018-6} {\bibfield  {journal} {\bibinfo  {journal}
  {Eur. Phys. J. A}\ }\textbf {\bibinfo {volume} {56}},\ \bibinfo {pages} {15}
  (\bibinfo {year} {2020})}\BibitemShut {NoStop}%
\bibitem [{\citenamefont {Sumiyoshi}\ \emph {et~al.}(2021)\citenamefont
  {Sumiyoshi}, \citenamefont {Fujibayashi}, \citenamefont {Sekiguchi},\ and\
  \citenamefont {Shibata}}]{sumiyoshi21}%
  \BibitemOpen
  \bibfield  {author} {\bibinfo {author} {\bibfnamefont {K.}~\bibnamefont
  {Sumiyoshi}}, \bibinfo {author} {\bibfnamefont {S.}~\bibnamefont
  {Fujibayashi}}, \bibinfo {author} {\bibfnamefont {Y.}~\bibnamefont
  {Sekiguchi}}, \ and\ \bibinfo {author} {\bibfnamefont {M.}~\bibnamefont
  {Shibata}},\ }\href {\doibase 10.3847/1538-4357/abce63} {\bibfield  {journal}
  {\bibinfo  {journal} {Astrophys. J.}\ }\textbf {\bibinfo {volume} {907}},\
  \bibinfo {pages} {92} (\bibinfo {year} {2021})}\BibitemShut {NoStop}%
\bibitem [{\citenamefont {Cusinato}\ \emph {et~al.}(2022)\citenamefont
  {Cusinato}, \citenamefont {Guercilena}, \citenamefont {Perego}, \citenamefont
  {Logoteta}, \citenamefont {Radice}, \citenamefont {Bernuzzi},\ and\
  \citenamefont {Ansoldi}}]{cusinato21}%
  \BibitemOpen
  \bibfield  {author} {\bibinfo {author} {\bibfnamefont {M.}~\bibnamefont
  {Cusinato}}, \bibinfo {author} {\bibfnamefont {F.~M.}\ \bibnamefont
  {Guercilena}}, \bibinfo {author} {\bibfnamefont {A.}~\bibnamefont {Perego}},
  \bibinfo {author} {\bibfnamefont {D.}~\bibnamefont {Logoteta}}, \bibinfo
  {author} {\bibfnamefont {D.}~\bibnamefont {Radice}}, \bibinfo {author}
  {\bibfnamefont {S.}~\bibnamefont {Bernuzzi}}, \ and\ \bibinfo {author}
  {\bibfnamefont {S.}~\bibnamefont {Ansoldi}},\ }\href {\doibase
  10.1140/epja/s10050-022-00743-5} {\bibfield  {journal} {\bibinfo  {journal}
  {Eur. Phys. J. A}\ }\textbf {\bibinfo {volume} {58}},\ \bibinfo {pages} {99}
  (\bibinfo {year} {2022})}\BibitemShut {NoStop}%
\bibitem [{\citenamefont {Sawyer}(1989)}]{sawyer89}%
  \BibitemOpen
  \bibfield  {author} {\bibinfo {author} {\bibfnamefont {R.~F.}\ \bibnamefont
  {Sawyer}},\ }\href {\doibase 10.1103/PhysRevC.40.865} {\bibfield  {journal}
  {\bibinfo  {journal} {Phys. Rev. C}\ }\textbf {\bibinfo {volume} {40}},\
  \bibinfo {pages} {865} (\bibinfo {year} {1989})}\BibitemShut {NoStop}%
\bibitem [{\citenamefont {{Reddy}}\ \emph {et~al.}(1998)\citenamefont
  {{Reddy}}, \citenamefont {{Prakash}},\ and\ \citenamefont
  {{Lattimer}}}]{reddy98}%
  \BibitemOpen
  \bibfield  {author} {\bibinfo {author} {\bibfnamefont {S.}~\bibnamefont
  {{Reddy}}}, \bibinfo {author} {\bibfnamefont {M.}~\bibnamefont {{Prakash}}},
  \ and\ \bibinfo {author} {\bibfnamefont {J.~M.}\ \bibnamefont {{Lattimer}}},\
  }\href {\doibase 10.1103/PhysRevD.58.013009} {\bibfield  {journal} {\bibinfo
  {journal} {Phys. Rev. C}\ }\textbf {\bibinfo {volume} {58}},\ \bibinfo
  {pages} {013009} (\bibinfo {year} {1998})}\BibitemShut {NoStop}%
\bibitem [{\citenamefont {Reddy}\ \emph {et~al.}(1999)\citenamefont {Reddy},
  \citenamefont {Prakash}, \citenamefont {Lattimer},\ and\ \citenamefont
  {Pons}}]{reddy99}%
  \BibitemOpen
  \bibfield  {author} {\bibinfo {author} {\bibfnamefont {S.}~\bibnamefont
  {Reddy}}, \bibinfo {author} {\bibfnamefont {M.}~\bibnamefont {Prakash}},
  \bibinfo {author} {\bibfnamefont {J.~M.}\ \bibnamefont {Lattimer}}, \ and\
  \bibinfo {author} {\bibfnamefont {J.~A.}\ \bibnamefont {Pons}},\ }\href
  {\doibase 10.1103/PhysRevC.59.2888} {\bibfield  {journal} {\bibinfo
  {journal} {Phys. Rev. C}\ }\textbf {\bibinfo {volume} {59}},\ \bibinfo
  {pages} {2888} (\bibinfo {year} {1999})}\BibitemShut {NoStop}%
\bibitem [{\citenamefont {{Mart{\'\i}nez-Pinedo}}\ \emph
  {et~al.}(2012)\citenamefont {{Mart{\'\i}nez-Pinedo}}, \citenamefont
  {{Fischer}}, \citenamefont {{Lohs}},\ and\ \citenamefont
  {{Huther}}}]{martinez-pinedo12}%
  \BibitemOpen
  \bibfield  {author} {\bibinfo {author} {\bibfnamefont {G.}~\bibnamefont
  {{Mart{\'\i}nez-Pinedo}}}, \bibinfo {author} {\bibfnamefont {T.}~\bibnamefont
  {{Fischer}}}, \bibinfo {author} {\bibfnamefont {A.}~\bibnamefont {{Lohs}}}, \
  and\ \bibinfo {author} {\bibfnamefont {L.}~\bibnamefont {{Huther}}},\ }\href
  {\doibase 10.1103/PhysRevLett.109.251104} {\bibfield  {journal} {\bibinfo
  {journal} {Phys. Rev. Lett.}\ }\textbf {\bibinfo {volume} {109}},\ \bibinfo
  {pages} {251104} (\bibinfo {year} {2012})}\BibitemShut {NoStop}%
\bibitem [{\citenamefont {Roberts}\ \emph {et~al.}(2012)\citenamefont
  {Roberts}, \citenamefont {Reddy},\ and\ \citenamefont {Shen}}]{roberts12}%
  \BibitemOpen
  \bibfield  {author} {\bibinfo {author} {\bibfnamefont {L.~F.}\ \bibnamefont
  {Roberts}}, \bibinfo {author} {\bibfnamefont {S.}~\bibnamefont {Reddy}}, \
  and\ \bibinfo {author} {\bibfnamefont {G.}~\bibnamefont {Shen}},\ }\href
  {\doibase 10.1103/PhysRevC.86.065803} {\bibfield  {journal} {\bibinfo
  {journal} {Phys. Rev. C}\ }\textbf {\bibinfo {volume} {86}},\ \bibinfo
  {pages} {065803} (\bibinfo {year} {2012})}\BibitemShut {NoStop}%
\bibitem [{\citenamefont {Pastore}\ \emph {et~al.}(2012)\citenamefont
  {Pastore}, \citenamefont {Martini}, \citenamefont {Buridon}, \citenamefont
  {Davesne}, \citenamefont {Bennaceur},\ and\ \citenamefont
  {Meyer}}]{pastore12}%
  \BibitemOpen
  \bibfield  {author} {\bibinfo {author} {\bibfnamefont {A.}~\bibnamefont
  {Pastore}}, \bibinfo {author} {\bibfnamefont {M.}~\bibnamefont {Martini}},
  \bibinfo {author} {\bibfnamefont {V.}~\bibnamefont {Buridon}}, \bibinfo
  {author} {\bibfnamefont {D.}~\bibnamefont {Davesne}}, \bibinfo {author}
  {\bibfnamefont {K.}~\bibnamefont {Bennaceur}}, \ and\ \bibinfo {author}
  {\bibfnamefont {J.}~\bibnamefont {Meyer}},\ }\href {\doibase
  10.1103/PhysRevC.86.044308} {\bibfield  {journal} {\bibinfo  {journal} {Phys.
  Rev. C}\ }\textbf {\bibinfo {volume} {86}},\ \bibinfo {pages} {044308}
  (\bibinfo {year} {2012})}\BibitemShut {NoStop}%
\bibitem [{\citenamefont {Iwamoto}\ and\ \citenamefont
  {Pethick}(1982)}]{iwamoto82}%
  \BibitemOpen
  \bibfield  {author} {\bibinfo {author} {\bibfnamefont {N.}~\bibnamefont
  {Iwamoto}}\ and\ \bibinfo {author} {\bibfnamefont {C.~J.}\ \bibnamefont
  {Pethick}},\ }\href {\doibase 10.1103/PhysRevD.25.313} {\bibfield  {journal}
  {\bibinfo  {journal} {Phys. Rev. D}\ }\textbf {\bibinfo {volume} {25}},\
  \bibinfo {pages} {313} (\bibinfo {year} {1982})}\BibitemShut {NoStop}%
\bibitem [{\citenamefont {Burrows}\ and\ \citenamefont
  {Sawyer}(1998)}]{burrows98}%
  \BibitemOpen
  \bibfield  {author} {\bibinfo {author} {\bibfnamefont {A.}~\bibnamefont
  {Burrows}}\ and\ \bibinfo {author} {\bibfnamefont {R.~F.}\ \bibnamefont
  {Sawyer}},\ }\href {\doibase 10.1103/PhysRevC.58.554} {\bibfield  {journal}
  {\bibinfo  {journal} {Phys. Rev. C}\ }\textbf {\bibinfo {volume} {58}},\
  \bibinfo {pages} {554} (\bibinfo {year} {1998})}\BibitemShut {NoStop}%
\bibitem [{\citenamefont {Horowitz}\ and\ \citenamefont
  {Schwenk}(2006)}]{horowitz06}%
  \BibitemOpen
  \bibfield  {author} {\bibinfo {author} {\bibfnamefont {C.~J.}\ \bibnamefont
  {Horowitz}}\ and\ \bibinfo {author} {\bibfnamefont {A.}~\bibnamefont
  {Schwenk}},\ }\href {\doibase 10.1016/j.physletb.2006.09.042} {\bibfield
  {journal} {\bibinfo  {journal} {Phys. Lett. B}\ }\textbf {\bibinfo {volume}
  {642}},\ \bibinfo {pages} {326} (\bibinfo {year} {2006})}\BibitemShut
  {NoStop}%
\bibitem [{\citenamefont {Horowitz}\ \emph {et~al.}(2017)\citenamefont
  {Horowitz}, \citenamefont {Caballero}, \citenamefont {Lin}, \citenamefont
  {O'Connor},\ and\ \citenamefont {Schwenk}}]{horowitz17}%
  \BibitemOpen
  \bibfield  {author} {\bibinfo {author} {\bibfnamefont {C.~J.}\ \bibnamefont
  {Horowitz}}, \bibinfo {author} {\bibfnamefont {O.~L.}\ \bibnamefont
  {Caballero}}, \bibinfo {author} {\bibfnamefont {Z.}~\bibnamefont {Lin}},
  \bibinfo {author} {\bibfnamefont {E.}~\bibnamefont {O'Connor}}, \ and\
  \bibinfo {author} {\bibfnamefont {A.}~\bibnamefont {Schwenk}},\ }\href
  {\doibase 10.1103/PhysRevC.95.025801} {\bibfield  {journal} {\bibinfo
  {journal} {Phys. Rev. C}\ }\textbf {\bibinfo {volume} {95}},\ \bibinfo
  {pages} {025801} (\bibinfo {year} {2017})}\BibitemShut {NoStop}%
\bibitem [{\citenamefont {Bedaque}\ \emph {et~al.}(2018)\citenamefont
  {Bedaque}, \citenamefont {Reddy}, \citenamefont {Sen},\ and\ \citenamefont
  {Warrington}}]{bedaque18}%
  \BibitemOpen
  \bibfield  {author} {\bibinfo {author} {\bibfnamefont {P.~F.}\ \bibnamefont
  {Bedaque}}, \bibinfo {author} {\bibfnamefont {S.}~\bibnamefont {Reddy}},
  \bibinfo {author} {\bibfnamefont {S.}~\bibnamefont {Sen}}, \ and\ \bibinfo
  {author} {\bibfnamefont {N.~C.}\ \bibnamefont {Warrington}},\ }\href
  {\doibase 10.1103/PhysRevC.98.015802} {\bibfield  {journal} {\bibinfo
  {journal} {Phys. Rev. C}\ }\textbf {\bibinfo {volume} {98}},\ \bibinfo
  {pages} {015802} (\bibinfo {year} {2018})}\BibitemShut {NoStop}%
\bibitem [{\citenamefont {Rrapaj}\ \emph {et~al.}(2015)\citenamefont {Rrapaj},
  \citenamefont {Holt}, \citenamefont {Bartl}, \citenamefont {Reddy},\ and\
  \citenamefont {Schwenk}}]{rrapaj15}%
  \BibitemOpen
  \bibfield  {author} {\bibinfo {author} {\bibfnamefont {E.}~\bibnamefont
  {Rrapaj}}, \bibinfo {author} {\bibfnamefont {J.~W.}\ \bibnamefont {Holt}},
  \bibinfo {author} {\bibfnamefont {A.}~\bibnamefont {Bartl}}, \bibinfo
  {author} {\bibfnamefont {S.}~\bibnamefont {Reddy}}, \ and\ \bibinfo {author}
  {\bibfnamefont {A.}~\bibnamefont {Schwenk}},\ }\href {\doibase
  10.1103/PhysRevC.91.035806} {\bibfield  {journal} {\bibinfo  {journal} {Phys.
  Rev. C}\ }\textbf {\bibinfo {volume} {91}},\ \bibinfo {pages} {035806}
  (\bibinfo {year} {2015})}\BibitemShut {NoStop}%
\bibitem [{\citenamefont {Bacca}\ \emph {et~al.}(2012)\citenamefont {Bacca},
  \citenamefont {Hally}, \citenamefont {Liebendorfer}, \citenamefont {Perego},
  \citenamefont {Pethick},\ and\ \citenamefont {Schwenk}}]{bacca12}%
  \BibitemOpen
  \bibfield  {author} {\bibinfo {author} {\bibfnamefont {S.}~\bibnamefont
  {Bacca}}, \bibinfo {author} {\bibfnamefont {K.}~\bibnamefont {Hally}},
  \bibinfo {author} {\bibfnamefont {M.}~\bibnamefont {Liebendorfer}}, \bibinfo
  {author} {\bibfnamefont {A.}~\bibnamefont {Perego}}, \bibinfo {author}
  {\bibfnamefont {C.~J.}\ \bibnamefont {Pethick}}, \ and\ \bibinfo {author}
  {\bibfnamefont {A.}~\bibnamefont {Schwenk}},\ }\href {\doibase
  10.1088/0004-637X/758/1/34} {\bibfield  {journal} {\bibinfo  {journal}
  {Astrophys. J.}\ }\textbf {\bibinfo {volume} {758}},\ \bibinfo {pages} {34}
  (\bibinfo {year} {2012})}\BibitemShut {NoStop}%
\bibitem [{\citenamefont {Bartl}\ \emph {et~al.}(2016)\citenamefont {Bartl},
  \citenamefont {Bollig}, \citenamefont {Janka},\ and\ \citenamefont
  {Schwenk}}]{bartl16}%
  \BibitemOpen
  \bibfield  {author} {\bibinfo {author} {\bibfnamefont {A.}~\bibnamefont
  {Bartl}}, \bibinfo {author} {\bibfnamefont {R.}~\bibnamefont {Bollig}},
  \bibinfo {author} {\bibfnamefont {H.-T.}\ \bibnamefont {Janka}}, \ and\
  \bibinfo {author} {\bibfnamefont {A.}~\bibnamefont {Schwenk}},\ }\href
  {\doibase 10.1103/PhysRevD.94.083009} {\bibfield  {journal} {\bibinfo
  {journal} {Phys. Rev. D}\ }\textbf {\bibinfo {volume} {94}},\ \bibinfo
  {pages} {083009} (\bibinfo {year} {2016})}\BibitemShut {NoStop}%
\bibitem [{\citenamefont {Hanhart}\ \emph {et~al.}(2001)\citenamefont
  {Hanhart}, \citenamefont {Phillips},\ and\ \citenamefont
  {Reddy}}]{Hanhart:2000ae}%
  \BibitemOpen
  \bibfield  {author} {\bibinfo {author} {\bibfnamefont {C.}~\bibnamefont
  {Hanhart}}, \bibinfo {author} {\bibfnamefont {D.~R.}\ \bibnamefont
  {Phillips}}, \ and\ \bibinfo {author} {\bibfnamefont {S.}~\bibnamefont
  {Reddy}},\ }\href {\doibase 10.1016/S0370-2693(00)01382-4} {\bibfield
  {journal} {\bibinfo  {journal} {Phys. Lett. B}\ }\textbf {\bibinfo {volume}
  {499}},\ \bibinfo {pages} {9} (\bibinfo {year} {2001})}\BibitemShut {NoStop}%
\bibitem [{\citenamefont {Lim}\ and\ \citenamefont {Holt}(2017)}]{lim17}%
  \BibitemOpen
  \bibfield  {author} {\bibinfo {author} {\bibfnamefont {Y.}~\bibnamefont
  {Lim}}\ and\ \bibinfo {author} {\bibfnamefont {J.~W.}\ \bibnamefont {Holt}},\
  }\href {\doibase 10.1103/PhysRevC.95.065805} {\bibfield  {journal} {\bibinfo
  {journal} {Phys. Rev. C}\ }\textbf {\bibinfo {volume} {95}},\ \bibinfo
  {pages} {065805} (\bibinfo {year} {2017})}\BibitemShut {NoStop}%
\bibitem [{\citenamefont {Carreau}\ \emph {et~al.}(2019)\citenamefont
  {Carreau}, \citenamefont {Gulminelli},\ and\ \citenamefont
  {Margueron}}]{carreau19}%
  \BibitemOpen
  \bibfield  {author} {\bibinfo {author} {\bibfnamefont {T.}~\bibnamefont
  {Carreau}}, \bibinfo {author} {\bibfnamefont {F.}~\bibnamefont {Gulminelli}},
  \ and\ \bibinfo {author} {\bibfnamefont {J.}~\bibnamefont {Margueron}},\
  }\href {\doibase 10.1140/epja/i2019-12884-1} {\bibfield  {journal} {\bibinfo
  {journal} {Eur. Phys. J. A}\ }\textbf {\bibinfo {volume} {55}},\ \bibinfo
  {pages} {188} (\bibinfo {year} {2019})}\BibitemShut {NoStop}%
\bibitem [{\citenamefont {Gandolfi}\ \emph {et~al.}(2011)\citenamefont
  {Gandolfi}, \citenamefont {Carlson},\ and\ \citenamefont
  {Pieper}}]{gandolfi11}%
  \BibitemOpen
  \bibfield  {author} {\bibinfo {author} {\bibfnamefont {S.}~\bibnamefont
  {Gandolfi}}, \bibinfo {author} {\bibfnamefont {J.}~\bibnamefont {Carlson}}, \
  and\ \bibinfo {author} {\bibfnamefont {S.~C.}\ \bibnamefont {Pieper}},\
  }\href {\doibase 10.1103/PhysRevLett.106.012501} {\bibfield  {journal}
  {\bibinfo  {journal} {Phys. Rev. Lett.}\ }\textbf {\bibinfo {volume} {106}},\
  \bibinfo {pages} {012501} (\bibinfo {year} {2011})}\BibitemShut {NoStop}%
\bibitem [{\citenamefont {Buraczynski}\ and\ \citenamefont
  {Gezerlis}(2016)}]{buraczynski16}%
  \BibitemOpen
  \bibfield  {author} {\bibinfo {author} {\bibfnamefont {M.}~\bibnamefont
  {Buraczynski}}\ and\ \bibinfo {author} {\bibfnamefont {A.}~\bibnamefont
  {Gezerlis}},\ }\href {\doibase 10.1103/PhysRevLett.116.152501} {\bibfield
  {journal} {\bibinfo  {journal} {Phys. Rev. Lett.}\ }\textbf {\bibinfo
  {volume} {116}},\ \bibinfo {pages} {152501} (\bibinfo {year}
  {2016})}\BibitemShut {NoStop}%
\bibitem [{\citenamefont {Baym}\ and\ \citenamefont {Kadanoff}(1961)}]{baym61}%
  \BibitemOpen
  \bibfield  {author} {\bibinfo {author} {\bibfnamefont {G.}~\bibnamefont
  {Baym}}\ and\ \bibinfo {author} {\bibfnamefont {L.~P.}\ \bibnamefont
  {Kadanoff}},\ }\href {\doibase 10.1103/PhysRev.124.287} {\bibfield  {journal}
  {\bibinfo  {journal} {Phys. Rev.}\ }\textbf {\bibinfo {volume} {124}},\
  \bibinfo {pages} {287} (\bibinfo {year} {1961})}\BibitemShut {NoStop}%
\bibitem [{\citenamefont {Harris}\ and\ \citenamefont
  {Evans}(1978)}]{harris78}%
  \BibitemOpen
  \bibfield  {author} {\bibinfo {author} {\bibfnamefont {C.~G.}\ \bibnamefont
  {Harris}}\ and\ \bibinfo {author} {\bibfnamefont {W.~A.~B.}\ \bibnamefont
  {Evans}},\ }\href {\doibase 10.1088/0022-3719/11/22/004} {\bibfield
  {journal} {\bibinfo  {journal} {J. Phys. C: Solid State Phys.}\ }\textbf
  {\bibinfo {volume} {11}},\ \bibinfo {pages} {4447} (\bibinfo {year}
  {1978})}\BibitemShut {NoStop}%
\bibitem [{\citenamefont {Entem}\ and\ \citenamefont
  {Machleidt}(2003)}]{entem03}%
  \BibitemOpen
  \bibfield  {author} {\bibinfo {author} {\bibfnamefont {D.~R.}\ \bibnamefont
  {Entem}}\ and\ \bibinfo {author} {\bibfnamefont {R.}~\bibnamefont
  {Machleidt}},\ }\href {\doibase 10.1103/PhysRevC.68.041001} {\bibfield
  {journal} {\bibinfo  {journal} {Phys. Rev. C}\ }\textbf {\bibinfo {volume}
  {68}},\ \bibinfo {pages} {041001} (\bibinfo {year} {2003})}\BibitemShut
  {NoStop}%
\bibitem [{\citenamefont {Coraggio}\ \emph {et~al.}(2007)\citenamefont
  {Coraggio}, \citenamefont {Covello}, \citenamefont {Gargano}, \citenamefont
  {Itaco}, \citenamefont {Entem}, \citenamefont {Kuo},\ and\ \citenamefont
  {Machleidt}}]{coraggio07}%
  \BibitemOpen
  \bibfield  {author} {\bibinfo {author} {\bibfnamefont {L.}~\bibnamefont
  {Coraggio}}, \bibinfo {author} {\bibfnamefont {A.}~\bibnamefont {Covello}},
  \bibinfo {author} {\bibfnamefont {A.}~\bibnamefont {Gargano}}, \bibinfo
  {author} {\bibfnamefont {N.}~\bibnamefont {Itaco}}, \bibinfo {author}
  {\bibfnamefont {D.~R.}\ \bibnamefont {Entem}}, \bibinfo {author}
  {\bibfnamefont {T.~T.~S.}\ \bibnamefont {Kuo}}, \ and\ \bibinfo {author}
  {\bibfnamefont {R.}~\bibnamefont {Machleidt}},\ }\href {\doibase
  10.1103/PhysRevC.75.024311} {\bibfield  {journal} {\bibinfo  {journal} {Phys.
  Rev. C}\ }\textbf {\bibinfo {volume} {75}},\ \bibinfo {pages} {024311}
  (\bibinfo {year} {2007})}\BibitemShut {NoStop}%
\bibitem [{\citenamefont {Marji}\ \emph {et~al.}(2013)\citenamefont {Marji},
  \citenamefont {Canul}, \citenamefont {MacPherson}, \citenamefont {Winzer},
  \citenamefont {Zeoli}, \citenamefont {Entem},\ and\ \citenamefont
  {Machleidt}}]{marji13}%
  \BibitemOpen
  \bibfield  {author} {\bibinfo {author} {\bibfnamefont {E.}~\bibnamefont
  {Marji}}, \bibinfo {author} {\bibfnamefont {A.}~\bibnamefont {Canul}},
  \bibinfo {author} {\bibfnamefont {Q.}~\bibnamefont {MacPherson}}, \bibinfo
  {author} {\bibfnamefont {R.}~\bibnamefont {Winzer}}, \bibinfo {author}
  {\bibfnamefont {C.}~\bibnamefont {Zeoli}}, \bibinfo {author} {\bibfnamefont
  {D.~R.}\ \bibnamefont {Entem}}, \ and\ \bibinfo {author} {\bibfnamefont
  {R.}~\bibnamefont {Machleidt}},\ }\href {\doibase 10.1103/PhysRevC.88.054002}
  {\bibfield  {journal} {\bibinfo  {journal} {Phys. Rev. C}\ }\textbf {\bibinfo
  {volume} {88}},\ \bibinfo {pages} {054002} (\bibinfo {year}
  {2013})}\BibitemShut {NoStop}%
\bibitem [{\citenamefont {Holt}\ and\ \citenamefont {Kaiser}(2017)}]{Holt}%
  \BibitemOpen
  \bibfield  {author} {\bibinfo {author} {\bibfnamefont {J.~W.}\ \bibnamefont
  {Holt}}\ and\ \bibinfo {author} {\bibfnamefont {N.}~\bibnamefont {Kaiser}},\
  }\href {\doibase 10.1103/PhysRevC.95.034326} {\bibfield  {journal} {\bibinfo
  {journal} {Phys. Rev. C}\ }\textbf {\bibinfo {volume} {95}},\ \bibinfo
  {pages} {034326} (\bibinfo {year} {2017})}\BibitemShut {NoStop}%
\bibitem [{\citenamefont {Wellenhofer}\ \emph {et~al.}(2014)\citenamefont
  {Wellenhofer}, \citenamefont {Holt}, \citenamefont {Kaiser},\ and\
  \citenamefont {Weise}}]{Wellenhofer:2014hya}%
  \BibitemOpen
  \bibfield  {author} {\bibinfo {author} {\bibfnamefont {C.}~\bibnamefont
  {Wellenhofer}}, \bibinfo {author} {\bibfnamefont {J.~W.}\ \bibnamefont
  {Holt}}, \bibinfo {author} {\bibfnamefont {N.}~\bibnamefont {Kaiser}}, \ and\
  \bibinfo {author} {\bibfnamefont {W.}~\bibnamefont {Weise}},\ }\href
  {\doibase 10.1103/PhysRevC.89.064009} {\bibfield  {journal} {\bibinfo
  {journal} {Phys. Rev. C}\ }\textbf {\bibinfo {volume} {89}},\ \bibinfo
  {pages} {064009} (\bibinfo {year} {2014})}\BibitemShut {NoStop}%
\bibitem [{\citenamefont {Whitehead}\ \emph {et~al.}(2021)\citenamefont
  {Whitehead}, \citenamefont {Lim},\ and\ \citenamefont {Holt}}]{Whitehead21}%
  \BibitemOpen
  \bibfield  {author} {\bibinfo {author} {\bibfnamefont {T.~R.}\ \bibnamefont
  {Whitehead}}, \bibinfo {author} {\bibfnamefont {Y.}~\bibnamefont {Lim}}, \
  and\ \bibinfo {author} {\bibfnamefont {J.~W.}\ \bibnamefont {Holt}},\ }\href
  {\doibase 10.1103/PhysRevLett.127.182502} {\bibfield  {journal} {\bibinfo
  {journal} {Phys. Rev. Lett.}\ }\textbf {\bibinfo {volume} {127}},\ \bibinfo
  {pages} {182502} (\bibinfo {year} {2021})}\BibitemShut {NoStop}%
\bibitem [{\citenamefont {Holt}\ \emph {et~al.}(2012)\citenamefont {Holt},
  \citenamefont {Kaiser},\ and\ \citenamefont {Weise}}]{Holt:2011yj}%
  \BibitemOpen
  \bibfield  {author} {\bibinfo {author} {\bibfnamefont {J.~W.}\ \bibnamefont
  {Holt}}, \bibinfo {author} {\bibfnamefont {N.}~\bibnamefont {Kaiser}}, \ and\
  \bibinfo {author} {\bibfnamefont {W.}~\bibnamefont {Weise}},\ }\href
  {\doibase 10.1016/j.nuclphysa.2011.12.001} {\bibfield  {journal} {\bibinfo
  {journal} {Nucl. Phys. A}\ }\textbf {\bibinfo {volume} {876}},\ \bibinfo
  {pages} {61} (\bibinfo {year} {2012})}\BibitemShut {NoStop}%
\bibitem [{\citenamefont {Holt}\ \emph {et~al.}(2013)\citenamefont {Holt},
  \citenamefont {Kaiser}, \citenamefont {Miller},\ and\ \citenamefont
  {Weise}}]{holt13}%
  \BibitemOpen
  \bibfield  {author} {\bibinfo {author} {\bibfnamefont {J.~W.}\ \bibnamefont
  {Holt}}, \bibinfo {author} {\bibfnamefont {N.}~\bibnamefont {Kaiser}},
  \bibinfo {author} {\bibfnamefont {G.~A.}\ \bibnamefont {Miller}}, \ and\
  \bibinfo {author} {\bibfnamefont {W.}~\bibnamefont {Weise}},\ }\href
  {\doibase 10.1103/PhysRevC.88.024614} {\bibfield  {journal} {\bibinfo
  {journal} {Phys. Rev. C}\ }\textbf {\bibinfo {volume} {88}},\ \bibinfo
  {pages} {024614} (\bibinfo {year} {2013})}\BibitemShut {NoStop}%
\bibitem [{\citenamefont {Holt}\ \emph {et~al.}(2016)\citenamefont {Holt},
  \citenamefont {Kaiser},\ and\ \citenamefont {Miller}}]{holt16}%
  \BibitemOpen
  \bibfield  {author} {\bibinfo {author} {\bibfnamefont {J.~W.}\ \bibnamefont
  {Holt}}, \bibinfo {author} {\bibfnamefont {N.}~\bibnamefont {Kaiser}}, \ and\
  \bibinfo {author} {\bibfnamefont {G.~A.}\ \bibnamefont {Miller}},\ }\href
  {\doibase 10.1103/PhysRevC.93.064603} {\bibfield  {journal} {\bibinfo
  {journal} {Phys. Rev. C}\ }\textbf {\bibinfo {volume} {93}},\ \bibinfo
  {pages} {064603} (\bibinfo {year} {2016})}\BibitemShut {NoStop}%
\bibitem [{\citenamefont {Rrapaj}\ \emph {et~al.}(2016)\citenamefont {Rrapaj},
  \citenamefont {Roggero},\ and\ \citenamefont {Holt}}]{rrapaj16}%
  \BibitemOpen
  \bibfield  {author} {\bibinfo {author} {\bibfnamefont {E.}~\bibnamefont
  {Rrapaj}}, \bibinfo {author} {\bibfnamefont {A.}~\bibnamefont {Roggero}}, \
  and\ \bibinfo {author} {\bibfnamefont {J.~W.}\ \bibnamefont {Holt}},\ }\href
  {\doibase 10.1103/PhysRevC.93.065801} {\bibfield  {journal} {\bibinfo
  {journal} {Phys. Rev. C}\ }\textbf {\bibinfo {volume} {93}},\ \bibinfo
  {pages} {065801} (\bibinfo {year} {2016})}\BibitemShut {NoStop}%
\bibitem [{\citenamefont {Qian}\ and\ \citenamefont
  {Woosley}(1996)}]{Qian1996}%
  \BibitemOpen
  \bibfield  {author} {\bibinfo {author} {\bibfnamefont {Y.-Z.}\ \bibnamefont
  {Qian}}\ and\ \bibinfo {author} {\bibfnamefont {S.~E.}\ \bibnamefont
  {Woosley}},\ }\href {\doibase 10.1086/177973} {\bibfield  {journal} {\bibinfo
   {journal} {Astrophys. J.}\ }\textbf {\bibinfo {volume} {471}},\ \bibinfo
  {pages} {331} (\bibinfo {year} {1996})}\BibitemShut {NoStop}%
\bibitem [{\citenamefont {Arcones}\ and\ \citenamefont
  {Thielemann}(2012)}]{Arcones2013}%
  \BibitemOpen
  \bibfield  {author} {\bibinfo {author} {\bibfnamefont {A.}~\bibnamefont
  {Arcones}}\ and\ \bibinfo {author} {\bibfnamefont {F.-K.}\ \bibnamefont
  {Thielemann}},\ }\href {\doibase 10.1088/0954-3899/40/1/013201} {\bibfield
  {journal} {\bibinfo  {journal} {J. Phys. G: Nucl. and Part. Phys.}\ }\textbf
  {\bibinfo {volume} {40}},\ \bibinfo {pages} {013201} (\bibinfo {year}
  {2012})}\BibitemShut {NoStop}%
\bibitem [{\citenamefont {Kajino}\ \emph {et~al.}(2019)\citenamefont {Kajino},
  \citenamefont {Aoki}, \citenamefont {Balantekin}, \citenamefont {Diehl},
  \citenamefont {Famiano},\ and\ \citenamefont {Mathews}}]{Kajino:2019abv}%
  \BibitemOpen
  \bibfield  {author} {\bibinfo {author} {\bibfnamefont {T.}~\bibnamefont
  {Kajino}}, \bibinfo {author} {\bibfnamefont {W.}~\bibnamefont {Aoki}},
  \bibinfo {author} {\bibfnamefont {A.~B.}\ \bibnamefont {Balantekin}},
  \bibinfo {author} {\bibfnamefont {R.}~\bibnamefont {Diehl}}, \bibinfo
  {author} {\bibfnamefont {M.~A.}\ \bibnamefont {Famiano}}, \ and\ \bibinfo
  {author} {\bibfnamefont {G.~J.}\ \bibnamefont {Mathews}},\ }\href {\doibase
  10.1016/j.ppnp.2019.02.008} {\bibfield  {journal} {\bibinfo  {journal} {Prog.
  Part. Nucl. Phys.}\ }\textbf {\bibinfo {volume} {107}},\ \bibinfo {pages}
  {109} (\bibinfo {year} {2019})}\BibitemShut {NoStop}%
\bibitem [{\citenamefont {Cowan}\ \emph {et~al.}(2021)\citenamefont {Cowan},
  \citenamefont {Sneden}, \citenamefont {Lawler}, \citenamefont {Aprahamian},
  \citenamefont {Wiescher}, \citenamefont {Langanke}, \citenamefont
  {Mart\'\i{}nez-Pinedo},\ and\ \citenamefont {Thielemann}}]{Cowan:2019pkx}%
  \BibitemOpen
  \bibfield  {author} {\bibinfo {author} {\bibfnamefont {J.~J.}\ \bibnamefont
  {Cowan}}, \bibinfo {author} {\bibfnamefont {C.}~\bibnamefont {Sneden}},
  \bibinfo {author} {\bibfnamefont {J.~E.}\ \bibnamefont {Lawler}}, \bibinfo
  {author} {\bibfnamefont {A.}~\bibnamefont {Aprahamian}}, \bibinfo {author}
  {\bibfnamefont {M.}~\bibnamefont {Wiescher}}, \bibinfo {author}
  {\bibfnamefont {K.}~\bibnamefont {Langanke}}, \bibinfo {author}
  {\bibfnamefont {G.}~\bibnamefont {Mart\'\i{}nez-Pinedo}}, \ and\ \bibinfo
  {author} {\bibfnamefont {F.-K.}\ \bibnamefont {Thielemann}},\ }\href
  {\doibase 10.1103/RevModPhys.93.015002} {\bibfield  {journal} {\bibinfo
  {journal} {Rev. Mod. Phys.}\ }\textbf {\bibinfo {volume} {93}},\ \bibinfo
  {pages} {15002} (\bibinfo {year} {2021})}\BibitemShut {NoStop}%
\bibitem [{\citenamefont {Chakraborty}\ \emph {et~al.}(2016)\citenamefont
  {Chakraborty}, \citenamefont {Hansen}, \citenamefont {Izaguirre},\ and\
  \citenamefont {Raffelt}}]{Chakraborty2016}%
  \BibitemOpen
  \bibfield  {author} {\bibinfo {author} {\bibfnamefont {S.}~\bibnamefont
  {Chakraborty}}, \bibinfo {author} {\bibfnamefont {R.}~\bibnamefont {Hansen}},
  \bibinfo {author} {\bibfnamefont {I.}~\bibnamefont {Izaguirre}}, \ and\
  \bibinfo {author} {\bibfnamefont {G.}~\bibnamefont {Raffelt}},\ }\href
  {\doibase https://doi.org/10.1016/j.nuclphysb.2016.02.012} {\bibfield
  {journal} {\bibinfo  {journal} {Nucl. Phys. B}\ }\textbf {\bibinfo {volume}
  {908}},\ \bibinfo {pages} {366} (\bibinfo {year} {2016})},\ \bibinfo {note}
  {neutrino Oscillations: Celebrating the Nobel Prize in Physics
  2015}\BibitemShut {NoStop}%
\bibitem [{\citenamefont {Tamborra}\ and\ \citenamefont
  {Shalgar}(2021)}]{Tamborra2021}%
  \BibitemOpen
  \bibfield  {author} {\bibinfo {author} {\bibfnamefont {I.}~\bibnamefont
  {Tamborra}}\ and\ \bibinfo {author} {\bibfnamefont {S.}~\bibnamefont
  {Shalgar}},\ }\href {\doibase 10.1146/annurev-nucl-102920-050505} {\bibfield
  {journal} {\bibinfo  {journal} {Ann. Rev. Nucl. Part. Sci.}\ }\textbf
  {\bibinfo {volume} {71}},\ \bibinfo {pages} {165} (\bibinfo {year}
  {2021})}\BibitemShut {NoStop}%
\bibitem [{\citenamefont {Vlasenko}\ \emph {et~al.}(2014)\citenamefont
  {Vlasenko}, \citenamefont {Fuller},\ and\ \citenamefont
  {Cirigliano}}]{Vlasenko2014}%
  \BibitemOpen
  \bibfield  {author} {\bibinfo {author} {\bibfnamefont {A.}~\bibnamefont
  {Vlasenko}}, \bibinfo {author} {\bibfnamefont {G.~M.}\ \bibnamefont
  {Fuller}}, \ and\ \bibinfo {author} {\bibfnamefont {V.}~\bibnamefont
  {Cirigliano}},\ }\href {\doibase 10.1103/PhysRevD.89.105004} {\bibfield
  {journal} {\bibinfo  {journal} {Phys. Rev. D}\ }\textbf {\bibinfo {volume}
  {89}},\ \bibinfo {pages} {105004} (\bibinfo {year} {2014})}\BibitemShut
  {NoStop}%
\bibitem [{\citenamefont {Richers}\ \emph {et~al.}(2019)\citenamefont
  {Richers}, \citenamefont {McLaughlin}, \citenamefont {Kneller},\ and\
  \citenamefont {Vlasenko}}]{Richers2019}%
  \BibitemOpen
  \bibfield  {author} {\bibinfo {author} {\bibfnamefont {S.~A.}\ \bibnamefont
  {Richers}}, \bibinfo {author} {\bibfnamefont {G.~C.}\ \bibnamefont
  {McLaughlin}}, \bibinfo {author} {\bibfnamefont {J.~P.}\ \bibnamefont
  {Kneller}}, \ and\ \bibinfo {author} {\bibfnamefont {A.}~\bibnamefont
  {Vlasenko}},\ }\href {\doibase 10.1103/PhysRevD.99.123014} {\bibfield
  {journal} {\bibinfo  {journal} {Phys. Rev. D}\ }\textbf {\bibinfo {volume}
  {99}},\ \bibinfo {pages} {123014} (\bibinfo {year} {2019})}\BibitemShut
  {NoStop}%
\bibitem [{\citenamefont {Johns}\ and\ \citenamefont
  {Xiong}(2022)}]{Johns2022}%
  \BibitemOpen
  \bibfield  {author} {\bibinfo {author} {\bibfnamefont {L.}~\bibnamefont
  {Johns}}\ and\ \bibinfo {author} {\bibfnamefont {Z.}~\bibnamefont {Xiong}},\
  }\href {\doibase 10.1103/PhysRevD.106.103029} {\bibfield  {journal} {\bibinfo
   {journal} {Phys. Rev. D}\ }\textbf {\bibinfo {volume} {106}},\ \bibinfo
  {pages} {103029} (\bibinfo {year} {2022})}\BibitemShut {NoStop}%
\bibitem [{\citenamefont {Johns}(2023)}]{Johns2023}%
  \BibitemOpen
  \bibfield  {author} {\bibinfo {author} {\bibfnamefont {L.}~\bibnamefont
  {Johns}},\ }\href {\doibase 10.1103/PhysRevLett.130.191001} {\bibfield
  {journal} {\bibinfo  {journal} {Phys. Rev. Lett.}\ }\textbf {\bibinfo
  {volume} {130}},\ \bibinfo {pages} {191001} (\bibinfo {year}
  {2023})}\BibitemShut {NoStop}%
\end{thebibliography}
%


\clearpage

\end{document}